# Diffuse Interstellar Bands: A Comprehensive Laboratory Study

Fred M. Johnson
*Cal State University Fullerton*
*Dept. of Physics, P.O .Box 6866, Fullerton, CA 92834, USA*

## Abstract

As a result of the search for the identity of the chromophores responsible for producing the diffuse interstellar bands, a comprehensive exposition of experimental data is presented, which implicates the following molecules: 1. The extremely stable organic molecules, magnesium tetrabenzoporphyrin (MgTBP) and $H_2$TBP. 2. A paraffin matrix (referred to as "grains") containing TBPs. 3. A low concentration of pyridine (also within the grains), whose transmission window at 2175Å, accounts for the ubiquitous UV bump. The blue emission spectra associated with the central star, HD44179, of the Red Rectangle displays the fluorescence excitation spectra of *bare* MgTBP. This unique spectrum matches the low temperature lab data of MgTBP in the vapor phase. An effective grain temperature of 2.728 K (+/- 0.008) was deduced, based on MgTBP's lowest measured vibrational state of 341 GHz.

## 1. Introduction

Since 1921, diffuse interstellar bands (DIBs) have presented an extreme challenge to both astronomers and spectroscopists. These bands are seen primarily in the spectroscopic data of distant, bright background stars, whose light has undergone multiple scattering associated with grains within intervening, interstellar clouds. The interstellar origin of DIBs was established by Paul Merrill [1]. At first, only a handful of these features were discovered, with one particular feature, the 4428Å DIB, being the most prominent. Several hundred DIBs are now being reported, a large number as yet unverified and unpublished. The quest for a solution to the DIB enigma has recently gained increased attention; its solution is regarded as potentially far-reaching in its implied cosmological significance.

G.H. Herbig [2] reviewed the historical background relating to the discovery of the DIBs, and discussed numerous proposals for their identification. The earliest summary of experimental work on the DIB enigma was presented in 1965, [3] concluding with the suggestion of complex hydrocarbon molecules as the carrier of the DIBs. Snow's review article [4], emphasized the experimental search for gas-phase molecules, overlooking critical solid-state literature, which includes high-quality Shpolskii spectral data [5-29].

The Shpolskii technique [8] is unique in that it facilitates the expression of well-resolved vibronic spectra of complex molecules such as porphyrins and phthalocyanines in frozen paraffin matrices. Moreover, Shpolskii spectral bandwidths are comparable to those of narrow DIBs, whose band widths are characterized as "diffuse" (in contrast to the sharp atomic spectral lines).

## 2. Proposed solution

The proposed solution to the DIB enigma utilizes the extensive and meticulously accumulated observations of Herbig [2, 30-36], survey data [37], and the Red Rectangle fluorescence emissions [38-39], plus 36 years of accumulated lab data of Magnesium-tetrabenzoporphyrin (MgTBP) (Fig. 1) and the free-base variety, $H_2$TBP [40-42]. These highly stable, abiotic molecules belong to a sub-class of a larger category of molecules known as, "porphyrins and metalloporphyrins" [43-45]. Porphyrins are extensively studied for their importance in biochemistry and biology. They have also held center stage in a large variety of photochemical studies where they serve as photosensitizers [46-51]. In fact, MgTBP has been used by Goedheer [52, 53] as a chlorophyll analogue.

Preliminary results have been given previously [5,6, 41-42, 54-63] (this paper supersedes, and amends all previous work on this subject). The two postulated DIB chromophores are within two special classes of organic compounds (grains) composed of the solvent pyridine and mixtures of paraffins. The primary assumption is that the main chromophores, MgTBP and $H_2$TBP, are situated within a set of grain types where the "A-type" produces the quasilines (via Shpolskii matrices) and the "B-type" produces the inhomogeneously broadened features. Overlapping vibronic transitions account for wider DIB line widths as well as for some of their enigmatic substructure line shapes. Factors such as, different experimental conditions and techniques, various equipment and technologies, and scientific contributions from others contributed to the 40-year quest for a unique solution.

## 3. Experimental techniques

A systematic spectroscopic survey of a large number of organic compounds [57] was performed on a Beckman DK9 spectrophotometer. Low-temperature absorption and fluorescence data collection utilized a ¾ m Jarrell Ash spectrometer (dispersion of 20Å/ mm) and was recorded on 5"x 7" photographic plates. Light sources included a high intensity tungsten lamp, HeNe, Argon Ion and/or HeCd (4416 Å) lasers.

MgTBP samples were prepared according to Linstead [64], Koehorst et al. [65] by various chemists over a ten-year period. One sample was a gift from Dr. Frank Träger. The Shpolskii matrix isolation method [8] was attempted throughout, with the tetrabenzoporphyrin samples first dissolved in pyridine and then diluted with either n-octane, n-nonane or n-decane.

Sample thickness was varied from about one to five millimeters. Absorption data required exposures of the order of minutes, whereas fluorescence data would vary from 15 minutes to 5 hours. A typical run would be seven separate exposures recorded on a single photographic plate. About two-dozen CSUF physics undergraduates collaborated in the data collection. Preliminary Shpolskii data reported previously [41, 42] plus over 120 spectral plates were meticulously re-measured by Perry Rose. Some of the plates were recently rescanned by David Weinkle, using modern technology, in order to facilitate an improved spectral representation of the data.

Room temperature FTIR spectra were taken of Mg TBP using both reflection and transmission set-ups. Raman spectra of MgTBP was obtained via Dr. Markwort.

The Q-band $S_1 \leftarrow S_0$ region of "bare" MgTBP was measured by Dr. Brumbaugh at the University of Chicago. The laser induced fluorescence (LIF) spectra covered the optical region from 5650Å to 6165Å. MgTBP was first heated to 425˚C to produce a vapor, which was subsequently cooled by means of a supersonic jet of helium.

Preliminary measurements of the $S_2 \leftarrow S_0$ transitions of "bare" MgTBP were taken by Dr. U. Even at Tel Aviv University. The experimental arrangement incorporates the following innovations:
(a) Pulsed supersonic gas expansion and simultaneous molecular ion mass determination.
(b) Molecular photo ionization is achieved by means of two synchronously pulsed laser beams, where one has a fixed and the other a variable frequency. Their combined energy was in slight excess of 7.3 e.v.

**4. Results**

The search originated with an earlier, unsuccessful experimental study at Columbia University involving F-centers in alkali hydrides [66].

Subsequent research involved absorption measurements of hundreds of diverse types of compounds, including a large variety of neutral polycyclic aromatic hydrocarbons (PAHs). The resulting spectra was inconsistent with the known DIB spectra [57]. In addition, an extensive library search provided the important clues, which led to the TBPs.

MgTBP is designated as the primary molecule. At room temperature [40] with MgTBP in a pyridine solution, the dominant spectral features are the Soret band at 4422Å, the Q band at 6344Å, the weaker Q (0, 1) broad structure at 5900Å and a vibronic structure at 4161Å. These features are far better resolved at temperatures of 77K and below, with slight temperature related spectral shifts. The peak of the 4422Å Soret shifts to 4428Å (and is weakly dependent on the pyridine/paraffin ratio); the 4161Å becomes 4175Å. The latter value is consistent with astronomical data by Herbig et al. [2, 67]. The Q (0, 1) and the Q bands show well-resolved structures, both in the lab and DIB data.

*4.1 Lab quasiline Shpolskii spectra at 77K*

The successful production of quasilines depends primarily on the paraffin solvents, on sample cooling rates and on the resulting microcrystalline structural environment in the vicinity of the MgTBP hosts. In addition, critical parameters include sample preparation, sample thickness, and the proper experimental techniques for producing and recording the spectra.

Mixtures were generally of arbitrary concentrations and, hence, resulted in a good diversity of absorption and fluorescence spectra. The Soret (4428Å) and Q-bands (about 6300Å), would generally be obtained as an inhomogeneous, broadened background [41, 42].

The following DIBs are isolated from the potential DIB list (on the basis of lab data, see section 4.5 ):

1. 6597Å, 6591Å, 6177Å, 4385Å, which are assigned to the free base derivative, $H_2$TBP .
2. 4882Å, which is tentatively assigned to a non-descript MgTBP fragment [5,6].

The remaining DIBs are tentatively assigned to MgTBP.

Shpolskii quasilines did not always materialize for MgTBP. At other times, different Shpolskii lines emerged depending on the "sites" that were generated upon cooling. A large amount of data collection was therefore needed in order to locate all possibilities. The accumulated Shpolskii lab data implies that on a purely random basis, a large variety of sites can arise [19-21, 24-26, 28, 68-71]. Possible variations include solvation, differing pyridine/paraffin ratios, different sets of paraffins, crystal structures, and chromophore configurations such as monomers, dimers, and with/without pyridine ligands attached to the central Mg atom.

Broad spectral features are indicative of the inhomogeneous case. Based on the 120+ lab experiments, at least five distinct sites associated with its lowest excited electronic state ($S_1$) occur most frequently. Figure 2 is a representation of the $S_1$ sites identified so far. The complete set of vibrational frequencies (Table 1), and the various identified electronic (0-0) origins of the $S_1$ and $S_2$ states, allow for decoding and identifying the majority of the reported DIBs.

Table 1 is a comprehensive listing of lab-measured vibrational frequencies of tetrabenzporphyrins plus magnesium porphyrin (MgP), measured under widely different physical conditions, and including a large amount of data accumulated from this work.
MgP is the basic porphyrin structure, viz. MgTBP minus its four attached benzene rings. The referenced [26] vibrations of MgP listed in Table 1, column 12, thus apply solely to the basic porphyrin ring.
With the exception of referenced data by Platenkamp and Canters [7] and that of Sevchenko et al. [11], all Shpolskii quasiline spectra of MgTBP and $H_2$TBP in this manuscript were generated in the author's laboratory. FTIR data was taken by the author in the CSUF chemistry lab (table 1, column 6).

Low temperature LIF spectra of bare MgTBP was commissioned (using the author's MgTBP samples) and taken by Dr. D. Brumbaugh (Fig. 4) and Dr. U. Even (Fig. 8a.). This data is displayed in columns 1 and 2 of Table 1a.. MgTBP Raman data was obtained from Dr. Lars Markwort and Dr. Neal Spingarn (Tables 1, column 13). Dr. Spingarn also kindly supplied the reflection data for MgTBP (Table 1c onward). The purpose of this additional (so far unpublished) outside data was to lend support to the overall internal self- consistency of the extended list of vibrational frequencies listed in Table 1.

The set of possible vibronic transitions in MgTBP: $S_1(\upsilon)\leftarrow S_0$, $S_2(\upsilon)\leftarrow S_0$, produce a fraction of the observed DIBs. The DIBs in the red region originate from transitions $S_0(\nu)\leftarrow S_1(0)$. The strongest lab (Soret) band at 4428Å (Fig. 3) is also the strongest DIB. Its huge (lab) molecular extinction coefficient (1.5 x $10^6$) allows for interstellar abundance determination. There are 2 x $10^{14}$ MgTBP molecules per $cm^2$ in this line of sight for HD183143.

Table 2 lists all the lab-measured Shpolskii bands of MgTBP and the corresponding possible and definitive DIBs. It displays an overwhelming number of matches to the precision of the hand-measured data (about +/- 1Å, consistent with spectral equipment resolution).

Tables 3 and 4 contain lab spectral data converted to wave numbers. Subsequent columns are the tentative, vibrational assignments compatible with entries from Table 1, connecting each transition to its respective origin, indicated at the top of each column. Each site has its own origins. For the $S_2$ state, the crystal field splitting is 35 $cm^{-1}$ and only one site has so far been identified. The origins were derived and confirmed from the totality of entries of all measured vibronic transitions.

An abbreviated list of lab absorption and fluorescence spectral data is displayed in Table 4, which shows entries for both fluorescence and absorption lines near the Q band origin, taken under different experimental conditions. Each entry of a lab line is given a tentative assignment of an electronic origin, site and corresponding vibrational frequency, derived from Table 1. The overall data supports the MgTBP identification, since it involves highly reliable vibrational frequencies, in addition to the actual DIB matches.(The reliability factor arises from the multiplicity of independent measurements having equal or almost equal values).

The columns in each table describe the following: "n" represents the number of separate individual measurements, where the wavelength values in column two are average values. The most prominent DIB at 4428Å [72] and the superimposed lab data produce an exact match, see Fig.3. Note, both the wavelength and "base" band width match (see discussion in section 4.4 of the Soret band width). Representative MgTBP Shpolskii spectra (apart from Figs 3-6) can be found in several publications [7, 11]. Fig. 4 shows a lab band at 6379 Å, which encompasses DIBs λλ 6379, 6376, 6362 and 6353. Additional representative Shpolskii spectral bands from this work are shown in Figures 5 and 6, where rescanned data from two and six separate plates, respectively, were superimposed.

*4.2 The $S_1 \leftarrow S_0$ transition (bare MgTBP)*

The data shown in Fig. 7 covers three different spectral regions, each having a different gain. Fig.7 shows the LIF spectra in the range 5650 to 6165 Å of Mg TBP seeded in a supersonic expansion of helium. The intense peak at 6115 Å is clearly the electronic origin (0-0) of the $S_1 \leftarrow S_0$ transition. The spectra displayed in Fig.7 are attributed to the electronic-vibrational excitations of the $S_1 \leftarrow S_0$ transitions, originating from the ground vibrational state of $S_0$. The 0-0 band asymmetry may originate from unresolved vibrational sequence congestion, involving low-frequency vibrational modes, which were not effectively cooled, based on analogous observations by Fitch [73] on phthalocyanines. Table 1 compares the data of MgTBP with that of ZnTBP [74].

The inset of Fig.7 shows in greater detail the region close to 0-0. The ultra-low frequency modes are enhanced in this scan, which was taken under conditions of increased gain.
.

*4.3 The $S_2 \leftarrow S_0$ transition (Soret region), of "bare" MgTBP*

The experiment confirmed the molecular weight of 532 for the MgTBP ion and coincidentally showed that MgTBP is stable against spontaneous UV photodisintegration in the 7.3 e.v. energy range.

Fig. 8a shows a preliminary excitation spectrum of "bare" MgTBP in the spectral region of 3900 to 3960 Å. The data is designated "preliminary" because ultimate cooling of seeded MgTBP molecules in the pulsed, high pressure HeNe mixture was not attained due to a valve assembly fatigue. Fig. 8a, nevertheless, shows sufficient resolution of peaks, (superimposed upon vibronic congestion) to permit the identification of the electronic origin of the $S_2 \leftarrow S_0$ transition as 3945.6 Å, based on the fact that only one other intense peak was found (on the long wavelength side) on scanning up to 4000Å. The 3947Å peak arises from a vibrational level 11 $cm^{-1}$ above $S_0$ (i.e. hot band) and thereby confirms the incomplete cooling in this experiment. Additional confirmation of this 11 $cm^{-1}$ excited state level is found in (a) the isolated MgTBP molecule LIF spectra (Fig.7) and (b) a direct microwave, room temperature absorption measurement by Dr. P. H. Siegel [75] of JPL, which gave 341 GHz. (Fig.9). The latter value was confirmed by measuring its second harmonic at about 680 Ghz .

*4.4* The *Soret bandwidth and the $S_2 \leftarrow S_0$ transitions of MgTBP in a frozen matrix*

Table 3 lists measured quasilines, plus the Soret band associated with the $S_2 \leftarrow S_0$ vibronic transitions of MgTBP in pyridine/paraffin matrices at 77K. The number of independent measurements (n) is indicated in column 1. The origins, $S_{2x}$ and $S_{2y}$ were determined to be 22,444 and 22,479 $cm^{-1}$ respectively. Additional confirmation of the 35 $cm^{-1}$ splitting is obtained by 25 vibrational assignments, including three pairs of quasilines, whose average energy separation is 34.3 $cm^{-1.}$

Note that the Soret band in Table 3 (shown schematically in Fig. 10) is not coincident with either of the electronic origins but 139 $cm^{-1}$ above the $S_{2x}$ state instead. It is postulated here that the unusually high intensity of the Soret band and its apparently irreducible vibronic bandwidth at 4K, (Fig. 8b) arises from the superposition of (possibly 17-36) vibronic transitions. This could comprise most of the measured vibrations from 11 to 247 $cm^{-1}$. There are 36 vibrations listed in Table 1a, which fall in this range.

The base band width of the important λ4428 DIB can be deduced from two independent results: (a) MgTBP, in a single crystal of n-octane at 4K [7] produced a Soret base band width of 279 $cm^{-1}$ (Fig. 8b), (b) the isolated MgTBP molecule scanned near its $S_2$ origin revealed a 244 $cm^{-1}$ wide base due to fortuitously insufficient cooling (Fig. 8a), thereby producing vibrational congestion. In the matrix, one has to add 35$cm^{-1}$ (crystal field splitting) to the measured 244 $cm^{-1}$ congested region to produce an effective base width (Δ) of 279 $cm^{-1}$. Therefore, both sets of results (a) and (b) are consistent with the DIB base band width of 51Å (279 $cm^{-1}$) for the λ4428 DIB [30, 76]. It also confirms the 139 $cm^{-1}$ displacement of the peak of the Soret band (to +/- 1 $cm^{-1}$) from the 0-0 origin (Fig 10).

Exact coincidence of both the peak wavelength of the λ4428.2 DIB and the MgTBP lab data, plus its base band width make a compelling case for the chromophore identification. Note, in particular, that the base band width is independent of interstellar scattering and possible DIB saturation. Further confirmation for the correct assignments for the double origins of the $S_2$ state (viz. 22444, 22479 $cm^{-1}$) can be seen in Table 5, where a large number of reported and probable DIBs in the blue spectral region are listed. The vibrational frequencies from Table 1, appropriate for each of the vibronic transitions are tabulated in columns 4 and 5. Note the special cases of wide band DIBs, λλ4595 and 4760. The listed, multiple, superimposed vibronic transitions reproduce both the correct observed band width *and* their *average values* match the measured DIB central peak wavelength.

New possible DIBs (tabulated in Table 5) were derived from measurements of unassigned (or incorrectly assigned) absorption bands in the quoted references. In the future, an additional criteria for DIB identification could be the method outlined above, whereby vibronic assignments (based on the code) could be the final determining factor.
Several identical vibrational frequencies are involved in the production of vibronic transitions from each of the origins $S_{2x}$ and $S_{2y}$ respectively. The three pairs in Table 5 are (4394, 4401Å), (4413, 4420Å) and (4381, 4525.5). Since vibrational frequencies in Table 1 are unique to the TBPs, their exact match in the vibronic transitions provides additional evidence for the MgTBP identification.

## 4.5 The $H_2$TBP molecule

### *4.51 The Soret Band of $H_2$TBP*

The lower extinction coefficient (relative to MgTBP) [40] of the Soret band in $H_2$TBP at λ4385 (Fig. 11) must be attributed to its lack of intrastate vibrational coupling. This would suggest that the central Mg atom plays an essential role in promoting (or facilitating) intrastate coupling, which, in turn, would produce the strongly enhanced Soret intensity in MgTBP. There is evidence for the presence of the relatively weak 4385 Å band in some of the published high S/N spectra [30].

### *4.52 The Q-Bands of $H_2$TBP*

Fig. 11 shows a non-Shpolskii absorption spectrum of $H_2$TBP at 77K. It displays the 4385 Å Soret band and the double Q bands. The band at 6592 Å corresponds to the probable DIB identified by Herbig [2] at 6591 Å. The other broader, irregular lab feature has an average value of 6177 Å. and would correspond to a broad DIB at that wavelength [2]. Fig. 7 by Johnson [42] displays a well-resolved vibronic spectrum of $H_2$TBP, establishing an unambiguous identification of one of the (0-0) origins at 6597 Å and providing an exact DIB match. Fig. 6 in Johnson [42] is an absorption spectrum of $H_2$TBP, displaying four of the multiple Shpolskii bands, previously referenced, as well as a band at 6174 Å. This band would correspond to the (29 Å FWHM) wide DIB at 6177 Å [2]. The identical $H_2$TBP spectrum (plate #201) is also shown in Fig. 4 of Johnson [41] (it was erroneously mislabeled as molecule χ). Thus, since the principle absorption bands of $H_2$TBP can be matched with their DIB counterparts, this molecule is assigned as one of the interstellar grain constituents.

*4.6 The UV Bump*

The 1964 discovery of a so-called UV bump at 2175Å [77] has been followed by a substantial amount of astronomical data [78-82] and parametric analyses [83]. However, no convincing experimental lab data had ever been previously presented or published. The important clue, which led to its ultimate identification in 1994 [6], was the correlation of this bump with the strength of the 4428Å DIB [84-85].

The 2175Å bump appears in the astronomical data as an absorption feature, whereas, it is, in fact, a transmission window between two very strong absorption bands of pyridine [86]. To demonstrate this experimentally at room temperature, a 10 cm long quartz cell filled with octane was used and a single drop of pyridine added. The result is shown in Fig.12. A typical UV bump superimposed on the lab data is shown in Fig.13.

Hence, the UV bump originates as a result of trapped energy (in the 2175Å band) within the interstellar grains, and becomes apparent as "missing energy" (absorption) as a result of multiple scattering processes. A fluffy interstellar grain, having pyridine content, would allow a partial penetration of incoming stellar radiation in the 2175Å band, which, unlike the case of a liquid medium, would not necessarily reemerge from the grain and thus be absorbed. Preliminary results of this experiment were reported previously [6]. Variability of UV bump bandwidths in different lines of sight, are directly related to the variability of pyridine concentrations in the grains.

*4.7 Fluorescence*

Prior to 1975, DIBs had never been seen in emission. There are now a significant number of examples of DIB emissions. These include two in some O-type stars [72] and at a minimum three towards HD29647 [84, 85], plus a large number of emission bands from the RR [87].

*4.71 The special case of HD29647*

The DIBs in HD29647 need further elaboration. As was pointed out previously [5], the unusually sharp DIBs (some of which are in emission) in this line of sight, are assigned to MgTBP molecules situated in matrices, which are essentially devoid of pyridine. Hence, they may be analyzed based on the work of Platenkamp and Canters [7], where these molecules were imbedded in a single crystal of n-octane at 4K, and their spectra recorded from a single site. Table 6 lists ten of these DIBs and their assigned vibronic transitions from origins 15962 and 15992 $cm^{-1}$, respectively. Fig. 14 is a copy of the original data [88] on which the characteristic 30-31 $cm^{-1}$ separation is shown, indicative of the $S_1$ crystal field splitting for site VI (Fig 2).

*4.8 The Red Rectangle (RR)*

The correspondence between RR emission features and DIBs, was pointed out by Fossey [89] and Sarre [90]. MgTBP lab data (Table 2) confirms 35 out of a possible 57 RR bands [39]. Based on observations by Scarrott et al. [91], it was revealed that RR emission profiles of λ5797Å, 5850Å and 6614Å change with distance from the central excitation source HD44179. Van Winckel et al. [87] provided a detailed analysis of these RR emission features. Fig. 15 (modified from Fig. 9, Van Winckel [87]) suggests that there is a ready explanation for the apparent discrepancy between the RR features and DIBs based on the measured vibrational spectra of MgTBP. It should be noted that Glinski and Anderson [92] also measured RR emission in the 5800Å region as a function of distance from the star and were, likewise, aware of the apparent RR wavelength offset from the 5797.11Å DIB.

*4.81 Resolution of the apparent RR vs. DIB discrepancy*

Note, in Fig. 15, that if one extrapolates the measured peak wavelengths as a function of distance to the star, the lower line (designated "A" for the 5797Å DIB) intersects at 5801Å. This corresponds to a difference in energy from the 5797.11Å DIB of 11.3 $cm^{-1}$. Similarly, for the lower panel, extrapolation to the star yields 5853.7 Å, which corresponds to an energy difference from the 5849.78Å DIB of 11.3 $cm^{-1}$. This value corresponds to the lowest measured vibrational frequency of MgTBP!!

The other extrapolation to the star yields 5818Å (line "B"). This corresponds to an energy difference of 62 $cm^{-1}$ (see Table 1a) from the 5797.11Å DIB.

For the lower panel of Fig. 15, extrapolations C, D and E intersect at the zero arc distance at 5856.2Å, 5859.3Å and 5861.3Å, respectively. This would correspond to vibrational frequencies, 18, 28 and 33 $cm^{-1}$ of MgTBP, (Table 1a).

Table 7 is an abbreviated extension of Table 4, listing potential "sites" and the corresponding frequencies (from Tables 1e and 1f) that are associated with the vibronic transitions for DIBs 5797.11 and 5849.78Å. The reason why the 62 $cm^{-1}$ vibration is favored for 5797Å may be related to the fact that three pairs of transitions have energy differences of 62 $cm^{-1}$. These pairs are: (1333 and 1271); (1400 and 1338) and (1437and 1375) $cm^{-1}$ respectively.

Similarly, the lower panel contains the following pairs of values: (1186,1153); (1212,1184) $cm^{-1}$, whose energy differences are 33 $cm^{-1}$ and 28 $cm^{-1}$, which correspond to the extrapolated intercepts E and D, respectively.

It is obvious that the closer the distance to the star, the greater the photon light intensity and, hence, a proportional increase in optical-pumping efficiency and an increased population of low lying vibrational levels.

*4.9 Discovery of hot, isolated (bare) MgTBP molecules in the RR nebula.*

The blue spectral emission features, extending from 3700 to about 4500A, had previously been attributed [93] to be scattered light from the central star HD44179. A re-examination for the purpose of this paper, resulted in an unexpected discovery. These features reveal an extended emission spectrum, which in fact, can be unambiguously assigned to the fluorescence excitation spectrum of bare MgTBP, displaying well-resolved vibrational excitations of the Soret band. The spectral feature with the greatest intensity is located at 3945.6 Å, which corresponds to the electronic origin (0-0) of the $S_2 \leftarrow S_0$ transition of MgTBP (Fig. 16 and lab data, Figs. 8a.).

The wide band width of these features implies vibronic congestion (see section 4.3) characteristic of a high temperature environment. These elevated temperatures in turn produce a Bolzmann distribution of excited vibrational levels of the $S_0$ ground state. The resulting so-called "hot bands," seen in the RR emission spectra, extend to about 4300 Å. Table 8 lists the wavelengths of all the blue emission features measured from the original RR spectra (Fig. 16, adapted from Fig. 1 in Schmidt et al [93]). The electronic $S_2$ state of MgTBP in the vapor phase is doubly degenerate. Hence, there is only a single 0-0 origin (25,344.7 $cm^{-1}$). A vibrational analysis is shown in Table 8 where the vibronic transitions originating from $S_2(0) \leftarrow S_0(\upsilon)$ and $S_0(\upsilon) \leftarrow S_2(0)$ are listed.

Note in Table 8, that a pair of vibrations (402 and 619) are symmetric about the 0-0 origin, with another set (401 and 617) in the hot band vibrational sequence. A total of 15 vibronic transitions are listed, whose frequencies are well established for MgTBP (see Table 1), thus, adding considerable weight to the MgTBP identification. Errors in measuring the blue RR emission bands are of the order of 1-2 Å. Hence, one has to allow for a +/- 3 $cm^{-1}$ range in assigning vibrational frequencies. The distribution of maximum peak intensities ($I_p$) of the hot bands can be used to determine an effective Boltzmann temperature for the MgTBP molecules. A plot of $\log_e (I_p)$ vs. excitation energy (Fig. 17) provides an effective excitation temperature of 1580 K. The probability that any other organic molecule (other than MgTBP, which sublimes) can tolerate and/or survive this apparently high temperature regime is very low.

## 5.0 Theory

The two chromophores MgTBP and $H_2TBP$ have spatial symmetries, which place them in point groups $D_{4h}$ and $D_{2h}$, respectively. For the former, Gouterman's [94] four-orbital model predicts a degenerate $S_1 \leftarrow S_0$ transition (Q-band) and a degenerate $S_2 \leftarrow S_0$ transition (Soret or B band), where both transitions are polarized in the molecular plane. The degeneracy of the excited states in MgTBP can be lifted by the action of a crystalline field that may originate within solids such as a Shpolskii host matrix or a single crystal.

The strong Soret and Q bands arise in part from the delocalized nature of the Π electrons shared by the 16 or 18 conjugated bonds in the ring. The Q band transition is made possible by configuration interaction mixing.

Since a large number of DIBs consist of overlapping vibronic transitions, the intensity (I) of each DIB resulting from such a combination is given by

$$I = \sum_i n_i m_i p_i q_i$$

Where,

n= number of crystal sites involved in the line of sight
m= transition probability for each vibronic transition
p= probability that a certain site is occupied

q=probability that a certain configuration (chromophore and lattice) is involved

An experimental measure of the p factor is obtained by the number of times a particular transition is observed in the lab spectra, and is indicated as such in the lab data.

A large number of DIB vibronic transitions involve identical vibrational frequencies. The latter implies that the nuclear potential surfaces in the $S_0$ and $S_1$ states are essentially identical, which is independently confirmed by various lab measurements on the TBPs.
The data reported here confirm some of the vibrational modes associated with the analogous zinc-tetrabenzoporphyrin, specifically related to the pyrrole structure, viz. 746-752 $cm^{-1}$ and 966-978 $cm^{-1}$, where the latter corresponds to the pyrrole breathing mode. The in-phase $C_aC_m$ stretching mode of $A_{lg}$ symmetry ($D_{4h}$ notation) reported for ZnTBP [95-96] at 1620 $cm^{-1}$ is also present in MgTBP. Other reported vibrations for ZnTBP are 1584,1520,1379,1335,1320 and 1235 $cm^{-1}$ have corresponding frequencies in MgTBP at 1585,1518,1374,1336,1321, and 1235 $cm^{-1}$ respectively, corresponding to the symmetric vibrations of the isoindole subunits.

The fact that both MgTBP and ZnTBP show vibrational modes that differ only by one or two wave numbers, is not unexpected, since the central metal plays only a very minor role in most of the fundamental frequencies of these tetrabenzoporphyrins. The pyrrole symmetric half ring stretching frequency is split into modes 1374 and 1335 $cm^{-1}$ for MgTBP, with the higher frequency mode localized primarily on the benzo rings. Bands 1132-1136 $cm^{-1}$ correspond to the $C_mH$ in-plane bending mode. Bands 1313-1321 $cm^{-1}$ correspond to the stretching frequency mode of $C_aN$. The 11.3 $cm^{-1}$ vibration is associated with the bending modes of the four outermost benzene rings [97].

Theoretical transition probabilities for fluorescence of vibronic bands, originating from a typical $S_1$ electronic state, are of course, quite different than transition probabilities for absorption from the ground state, $S_0$. This fact can be readily seen in the experimental data [7]. To date, unidentified infrared (UIR) emission bands are routinely being compared with e.g. PAH molecular absorption data, rather than the appropriate fluorescence transition probabilities, or fluorescence lab data.

**6.0. The Interstellar Grain Temperature**

Assume that the lowest vibrational energy of MgTBP ($\nu$ = 341 GHz) (See Figure 9) is transferred to the phonon vibrational modes of its host crystal lattice (paraffin matrix). The laws of thermodynamics [98] provide the following equation:

$$\frac{1}{2}NkT = h\nu$$

where $N$ is the number of degrees of freedom of the four ($n$) outermost benzene ring bending modes, plus the 2 pyridine molecules attached to the central Mg atom, giving a total value for $n = 6$. The Boltzmann and Planck's constants, $k$ and $h$ respectively, are well - known. Thus, $N = 3n - 6$ and since $n = 6$; $N = 12$.

Substituting all known and measured values, gives

$$T = 2.728\,\mathrm{K}\ (\pm 0.008)$$

This implies that grains containing MgTBP would essentially radiate as black bodies at an equivalent temperature of 2.728 K. This is the temperature attributed to the cosmic background radiation, and hence has the potential to negate the Big Bang Theory.

## 7.0 Discussion

Weak DIBs seen as substructures by Krelowski and Sneden [99] and others are not only consistent with the proposed solution outlined in this paper, but seem to verify it in detail.

### *7.1 DIB Absorption vs. Emission*

The processes by which the DIBs are generated in the interstellar medium (ISM) have to be reexamined. We are dealing primarily with a frequency sensitive multiple scattering phenomena. Based on the UV bump identification and the observation of isolated ISM DIB emission features, rather than absorption, provides the important forensic clue that one has to consider in detail the scattering process for each grain. Frozen paraffins have a white, fluffy, snow-like appearance, hence incident radiation may simply scatter or else has an opportunity to get trapped. Fluorescence or phosphorescence radiation generated within such fluffy cocoon structured grain material will be substantially trapped. Likewise, resonant radiation arising from the $S_1$ state will also be trapped, thus in effect lengthening the $S_1$ lifetime. Subsequent radiation of frequency $\upsilon_s = [S_1(0)\text{-}S_0(\upsilon)]/h$ would stimulate the emission of a photon from the $S_1$ state. Consequently, incoming radiation at frequency $\upsilon_s$ would, likewise be absorbed and be represented in the DIBs spectra as a missing (absorption) band in the multi-scattering process.

Transitions of the type $S_0(\upsilon) \leftarrow S_1(0)$ are generally seen in the ISM as absorption, however they are observed as fluorescence in the RR. Therefore, one may deduce that some grains in the RR are small (and probably young), and did not as yet coalesce to form the relatively larger, fluffy grain aggregates, which are presumed to be predominantly prevalent in the ISM. Fluffy grain particles [100] had been analyzed from the point of view of scattering theory, but not for their relevant implications to interstellar spectroscopy.

### *7.2 The DIB-grain Connection*

In general, DIB strength, according to Herbig [34] is related (with some exceptions) to color excess, which implies a close association with grains. There are, however, an additional six, more cogent reasons why DIB chromophores have to be within grains.

1. Line widths consideration: in supersonically cooled, “bare” molecules [74] and this paper (Fig.7), vibronic lines have widths of the order of 1 $cm^{-1}$, which are extremely narrow, and therefore are inconsistent with the observed DIB line widths. Free molecules in interstellar space cannot readily produce, with precision, the wide range of observed DIB bandwidths despite the many valiant and unsuccessful attempts by Herzberg [101-103] and his successors. Note: Intrinsic line widths of interstellar atomic and diatomic spectral lines are invariably sharp and narrow, and never diffuse.

2. In special Shpolskii [8] type matrices, narrow lines are indeed produced, which are comparable in width to those observed in some of the narrow DIBs [2]. Wider DIB lines arise from overlapping vibronic transitions (see examples in Table 5) or inhomogeneously broadened lines. Use of the “code” (i.e. Tables 1+ and Fig.2) would extend this number to encompass virtually all of the DIBs.

3. Free molecules cannot readily account for specific transitions of DIB absorption in the interstellar medium and their appearance as fluorescence in the Red Rectangle spectra.

4. There are 15 pairs of DIBs [36] (and possibly more), which display the characteristic, crystal field-induced splitting (whose average value is 35.6 $cm^{-1}$) of the otherwise degenerate $S_1$ electronic state. This is consistent with MgTBP matrix lab data, and reconfirms the “inside grain” hypothesis.
5. Among the multitude of 35-36 $cm^{-1}$ wide DIB pairs, is the important Q-band duo of site I, viz. λλ6284 and 6270 (Fig. 5 and Table 4).
6. A total of well over 100 DIBs have so far been matched with the lab Shpolskii data to +/- 1Å [see Table 2]. The Shpolskii matrices would correspond to the solid state grains.

### *7.3 An Unintended Confirmation*

Sevchenko [11] produced a set of (Shpolskii) quasilines of MgTBP in a pyridine-acetone-paraffin matrix at 77 K. Of the 14 lines shown in his spectra, six match the DIBs to a precision of 1Å. With the aid of the “code”, i.e. Tables 1+ and knowledge of the $S_1$ sites (Fig.2), each of his lines belong to a single site $S_1$ (15,918 $cm^{-1}$). It should be pointed out that Sevchenko [11] was unaware of both the crystal field splitting and the relationship to the DIB problem. Moreover, Sevchenko did not explore the Soret region of MgTBP, hence the absence of the 4428Å band in his spectra.

### *7.4 Implications and Related Issues*

Johnson, Bailey and Wegner [58] considered the interstellar synthesis of MgTBP. Cosmochemical evolution of large organic molecules was analyzed and addressed by Hodgson [104-108], with particular emphasis on porphyrins, using plasma synthesis technology.

Containment within a grain matrix retards the TBP destruction in the harsh interstellar environment [109]. However, its ultimate demise does lead to about 19 possible molecular fragments [59] already identified by microwave techniques [110-111]. Among these relevant, lab-measured fragments, is the unique interstellar molecule, MgNC [112]. Thus, the origin of this enigmatic radical is no longer a mystery.

Implications of this identification were considered: vis a vis cosmology and the origin of life [113, 114]. It might also be of astronomical interest that MgTBP molecules are paramagnetic in their $S_1$ singlet state [7, 29] and thus may contribute to (or be the source of) the weak postulated interstellar magnetic fields.

Extensive searches for *polarization* of DIB signatures have so far proved illusionary. This is consistent with the hypothesis that multiple scattering within the fluffy grain cocoon would vitiate any such chromophore signals. A substantial amount of high-quality Shpolskii spectral data (see the extensive list of cited references, originating largely from the Russian and Dutch workers) is clearly at variance with theoretical predictions [115-116] that profile variations, such as emission wings, are to be expected if the DIB features were created by impurity absorbers in grains. These theoretical predictions, which apparently do not apply to Shpolskii matrices, have unfortunately proven a major stumbling block to progress in this field.

**8.0 Conclusions**

The five characteristics that distinguish MgTBP and $H_2$TBP from other organic compounds are:

(a) Thermodynamic stability (they sublime at 500˚ C).
(b) Unique spectra (both wide and narrow lines depending on matrix).
(c) MgTBP's Soret band, at 4428Å, has an extremely large molar extinction coefficient ($\sigma = 1.5 \times 10^6$).
(d) Approximately 200 vibrations of MgTBP (which account for its stability), plus the doublets of five or six potential $S_1$ electronic "sites", are capable to produce a large number of potential vibronic transitions. This would result in a plethora of possible DIBs, predominately located in the red spectral region.
(e) MgTBP is a highly efficient emitter of fluorescence radiation, specifically in the red spectral region, with photo-excitation in the blue spectral region.

An effective grain temperature was deduced from MgTBP's measured lowest vibrational state of 341 GHz: it is 2.728 K (+/- 0.008). This implies that grains containing MgTBP would essentially radiate as black bodies at an equivalent temperature of 2.728 K. This is the temperature attributed to the cosmic background radiation, and hence has the potential to negate the Big Bang Theory.

The combined weight of all the data collected in this paper points to one simple and unique conclusion. The success of critical scholarship and a diversity of relevant spectroscopic research is a quintessential demonstration of the compelling quality of Occam's razor.

**Acknowledgement**

Among the many outstanding CSUF physics students who participated in the spectroscopy were John Anthes, John Kerns, Ph.D., Mr. W. Krone-Schmidt, Mrs. L. Krone-Schmidt, Perry Rose, Anne Arroyo, Roberto Razo and Graciella Flores. David Weinkle took on the arduous task of rescanning Shpolskii data of a large number of photographic plates, utilizing his upgrades of a Jarrell Ash scanning densitomer. The generous scientific con- tributions of the following scientists are hereby most gratefully acknowledged: Dr. Don Brumbaugh, Dr. Benny Ehrenberg, Dr. Uzi Even, Dr. George Herbig, Dr. Lars Markwort, Dr. Franz- Peter Montforts, Dr. Roelof Platenkamp and Dr. Frank Träger. I wish to thank, in particular, Dr. Peter Siegel of JPL for pro- viding the important mm microwave absorption measurement of MgTBP. Thanks as well to Dr. Neil Spingarn for supplementing Table 1 with new Raman and Reflection data. Helpful suggestions by the anonymous referee are appreciated and resulted in improving this manuscript.

# LIST OF TABLES

**Table 1a**. Vibrational Frequencies of Tetrabenzoporphyrins plus Magnesium Porphyrin (MgP).

| MgTBP | | | | | | | CdTBP | ZnTBP | | MgP | $H_2$TBP | MgTBP |
|---|---|---|---|---|---|---|---|---|---|---|---|---|
| | | [7] | | Johnson | | | [7] | [74] | [13] | [26] | Johnson | |
| Bare | | Matrix, n-Octane 4 K | | Matrix, Quasiline 77 K | | FTIR | 4 K | Bare | Matrix | Matrix, 4 K | Matrix, 77 K | Raman |
| $S_1$ | $S_2$ | $S_1$ | $S_2$ | $S_0$ | $S_2$ | $S_0$ | *(S)* | *(S)* | *(S)* | *(S)* | *(S)* | $S_0$ |
| 1 | 2 | 3 | 4 | 5 | 6 | 7 | 8 | 9 | 10 | 11 | 12 | 13 |
| 10 | 11 | | 10.5 | 11 | | 11.3 microwave | | 10 a | | | | |
| | 18 | 17 | 18 | 18 | | | | 18 a | 18 | 17 | | |
| 20 | | | 20 | 20 | 21 | | | | | 21 | | |
| | | | 22.5 | 22 | | | | | | | | |
| 24 | | | 25 | 25 | | | | | 24 | | | |
| | 26 | | 27 | 27 | 28 | | | | | 27 | | |
| | 33 | | 33 | | | | | | | 35 | | |
| 40 | 40 | | 40 | 41 | 42 | | | | | | | |
| 44 | 44 | | 44 | 44 | | | 44 | | | 43 | | |
| | 61 | 62 | 61.5 | 62 | | | 62 | | | 62 | | |
| | 74 | | 73 | 73 | | | | | | | | |
| 77 | 78 | | 77 | 77 | | | | | 80 | | | |
| | 85 | | 85 | 85 | 85 | | | | | 87 | | |
| 94 | 92 | 90 | 92 | 92, 93 | 93 | Start of FTIR data | | | | | 90 | |
| | 100 | | 99 | 98 | | | | | | | 100 | |
| 110 | 110 | 113 | 109/111 | 111 | | 111 | | | | | | |
| 118 | 117 | 119 | 117 | 116, 118 | | | | | | | | |
| 128 | 126 | | 126 | 125 | | | | 124 | 123 | | | |
| 131 | | | | 131 | | 129 | | | 131 | | | |
| 146 | | | 146 | 146 | | 143 | | | | | | |
| | 149 | | 150 | 152 | | 152 | 150 | 148 | | | | |
| 156 | 155 | | 156 | 154, 156 | 156 | | | | | | | |
| | | 159, 162 | 161 | 159, 161 | | 161 | | | | | | |
| | 168 | | 167 | | | | | | | | | |
| 172 | | 172, 174 | 172 | 173, ‹175› | | | | | | | | |
| 172 | | 172, 174 | 172 | 173, ‹175› | | | | | | | | |
| | | | 180 | 180 | | 181 | | 179 | | | 182 | |
| | 187 | | 186 | 185 | | | | | | | | |
| | | | 189 | | | | 189, 191 | | | 189 | 190 | |
| | 197 | 194 | 195 | 195, 197 | | 195 | 197 | | 197 | 193 | | |
| | 210 | | 209 | 210 | | | | | | | | |
| 217 | | | 215 | 216 | | | | | | | 217 | |
| | 220 | 221 | 221 | 219, 221 | | 220 | | | | 220 | 220 | |
| 231 | | | 230 | 230 | | | | | 234 | 227 | 227 | |
| 238 | 238 | 239 | vs 238 | 238 | | | 238 | 238 | | 236 | | 239 |
| | | | 242 | 242 | | | | 242 | | | | |
| | 244 | 244 | vs 244 | 244 | | | | | 243 | | | |
| 247 | | 248 | 246 | 247 | | | | 248 | | | 247 | |
| | End of data | 257 | 257 | | | 255 | 256, 258 | 254 | | | | |
| | | 261 | 260 | 259 | | | 259 | 257 | | | | |
| | | | 265 | | | 264 | | | 264 | | | |
| | | | 270 | | | 270 | | | | | | |
| | | 276 | 275 | 275 | | | | | | | | |
| | | | 282 | 281 | | | | 281 | | | | |
| | | 287 | 287 | 288 | | 287 | | 288 | | | | |
| | | | 290 | | | | | | | | | |
| | | 303 | 297 | 304 | | | | 301 | | | | |
| 312 | | 311 | 311 | 311 | | 311 | 309 | | | | | |

Columns 1 and 2 contain data derived from spectra originally taken for this project by Drs. D. Brumbaugh and U. Even, respectively.

**Table 1b.** Vibrational Frequencies of Tetrabenzoporphyrins plus Magnesium Porphyrin (MgP)

| MgTBP | | | | | | CdTBP | ZnTBP | | $H_2$TBP | MgP | MgTBP |
|---|---|---|---|---|---|---|---|---|---|---|---|
| | [7] | | Johnson | | | [7] | [74] | [13] | Johnson | [26] | |
| Bare | *n*-Octane 4 K | | Quasiline 77 K | | FTIR | 4 K | Bare | Matrix | Matrix | 4 K | Raman |
| $S_1$ | $S_1$ | $S_0$ | $S_0$ | $S_1$ | $S_0$ | $S_0$ | $S_1$, $S_2$ | *(S)* | *(S)* | *(S)* | $S_0$ |
| 1 | 2 | 3 | 4 | 5 | 6 | 7 | 8 | 9 | 10 | 11 | 12 |
| | | | | | 334 | | | | 328 | | 330 w |
| | 341 | | 342 | | | | 351 | | | | |
| | 353 | | | | 355 | | 355 | | | 365 | |
| | 359 | | 357, 360 | | 363 | | 362 | | | | |
| 378 | | | 378 | | 376 | | 366 | | | | |
| 386 | 383 | | 383, 388 | | | | 386 | 386 | | | |
| 393 | | | 392, 394 | | 390 | | | | | 394 | |
| 401 | | | 400 | | | | 398 | 397 | 402 | 401 | |
| | | | 442 | | 422 | | | | 431 | | |
| | 460 | 462 | 461 | | | | 456, 460 | | 452 | | |
| | | | | | 474 | | 468 | | | | 475 |
| 478 | 477 | 480 | 478, 481 | | 480 | 477, 480 | 480 | 482 | | | 477, 480 |
| | | | 485 | | | | | 486 | | | |
| 488 | 489 | 487 | 489, 491 | | | | 490 | | | | |
| 505 | 504 | | 502, 504 | 502 | 501 | | 498 | 499 | 499, 503 | | |
| 507 | 507 | | 507 | 507 | 507 | | | | | | |
| 514 | 519 | | 518 | | | | 515 | 512 | 517 | | |
| | | | | | 521 | | | | 522 | | |
| | 539 | | 539 | 538 | 532 | 538 | 540 | | | | |
| 548 | 548 | | 547, 549 | | | | 548 | | | | |
| | | | 552 | | 550 | | | | | | 551 |
| 557 | 558, 562 | | 559 | 557 | | | 559 | | | | |

Commissioned LIF data of MgTBP by Dr. Brumbaugh column (1) and Raman data (column 12) supplied by Drs. Markwort and N. Spingarn.

**Table 1c** Vibrational Frequencies of Tetrabenzoporphyrins plus Magnesium Porphyrin (MgP)

| MgTBP | | | | | | | CdTBP | ZnTBP | | $H_2$TBP | MgP | MgTBP |
|---|---|---|---|---|---|---|---|---|---|---|---|---|
| | [7] | | Johnson | | | Spingarn | [7] | [74] | [13] | Johnson | [26] | |
| Bare | *n*-Octane 4 K | | Quasiline 77 K | | FTIR | Reflection 300 K | 4 K | Bare | Matrix | Matrix | 4 K | Raman |
| $S_1$ | $S_1$ | $S_0$ | $S_0$ | $S_1$ | $S_0$ | $S_0$ | $S_0$ | $S_1$, $S_2$ | $(S)$ | $(S)$ | $(S)$ | $S_0$ |
| 1 | 2 | 3 | 4 | 5 | 6 | 7 | 8 | 9 | 10 | 11 | 12 | 13 |
| 566 | 569 | | 569 | 568 | 571 | | | | 568 | 566 | | |
| 577 | 578 | | 580 | | 580 | | 578 | 577 | 575 | | | 580,581 |
| | | 582 | 583, 585 | | 584 | | | | | | | 583 |
| 591 | 591/592 | 593 | 593 | | | | | 589 | | 588 | | |
| | | | 595 s | 593 | 594 | | | | | | | |
| 599 | 599.5 | | 599 | 597 | 600 | | 602 | | 601 | 598 | | 598 w |
| | 615 | | 616 | 615 | 615 | | | | | | | 618 |
| 619 | 621 | 625 | 626 | 628 | 625 | | | | | | | |
| 629 | 629 | 629 | | | 633 | | | | | | | |
| 635 | 639 | 637 | 636, 639 | 639 | 640 w | | | | 633 | | | 637 |
| | 646 | | 646 | 643, 646 | 650 | | 651 | | | | | |
| | 659 | 659 | 659 | 659 | 660 | | | | | 660 | | |
| 675 | 678 | 679, 674 | 675 | 676 | 679 w | | 676 | | | 674 | | |
| 689 | | | | | | | 686 | | | | | |
| 692 | 691 | | | | | | | 691 | | 693 | | |
| 695 | | 697 | 695 | 695 s | 695 s | | | 698 | | 697 | | 696 s |
| 701 | | 704 | 702 | 702 s | 705 | | | 704 | 703 | 701 | | 701 |
| 710 | 709 | 707 | 708 | 706 | 708 | 707 | 708 | | 710 | | 708 | |
| 713 | | 712 | 711, 712 | 711 | | | | | | | 713 | |
| | | 722 | 721, 725 | | | | | | | 720 | 724 | |
| 728 | | 727 | 727 | | | | | 729 | | 726 | 730 | |
| 733 | | 732 | 732, 734 | 731 | | | | | | | 734 | 733, 734 |
| 736 | | 737 | 737 | 737 | | 738 | | | | | | 737 |
| 739 | 739 | 739 | 742 | | 741 | | | 739 | 740 | | 741 | 739 |
| 751 | | 752 | 751 | | | 754 | | | 745 | | | |
| 760 | | 758 | | | 757 | | | | | | | |
| 763 | | 762 | 763 | | 764 | | | 764 | | | | 760 sh |

Commissioned LIF data of MgTBP by Dr. Brumbaugh column (1) and Raman data (column 13) supplied by Drs. Markwort and N. Spingarn.

**Table 1d** Vibrational Frequencies of Tetrabenzoporphyrins plus Magnesium Porphyrin (MgP)

| MgTBP | | | | | | | CdTBP | ZnTBP | | $H_2$TBP | MgP | MgTBP |
|---|---|---|---|---|---|---|---|---|---|---|---|---|
| | [7] | | Johnson | | | Spingarn | [7] | [74] | [13] | Johnson | [26] | |
| Bare | *n*-Octane 4 K | | Quasiline 77 K | | FTIR | Reflection 300 K | 4 K | Bare | Matrix | Matrix | 4 K | Raman |
| $S_1$ | $S_1$ | $S_0$ | $S_0$ | $S_1$ | $S_0$ | $S_0$ | $S_0$ | $S_1, S_2$ | $(S)$ | $(S)$ | $(S)$ | $S_0$ |
| 1 | 2 | 3 | 4 | 5 | 6 | 7 | 8 | 9 | 10 | 11 | 12 | 13 |
| 769 | | 769 vs | 768, 770 | 767 | | | | | | | | |
| 776 | | 774 | 774 | 773 | | | | | | | | |
| 783 | 783 | | | | | | | 784 | | | 782 | 784 |
| 786 | | 788 | 787 | 787 s | | | | | | 786 | | |
| 792 | | | 794 | | | | | 792 | | | | |
| | | | | | | | | | | 798, 801 | 798 | |
| 818 | | 818 | 819 | | | | | 816 | | | | 818 s |
| | – | 821 | | | | | 820 | 820 | | | 821 | 820 |
| 824 | 825 | 826 | 824 | | | | 825, 828 | 827 | 827 | | 824 | 825 |
| 830 | 832 | 833 | 831 | | 831 | 830 | 830 | | | 830 | 827 | |
| 835 | | – | 836 | | | | – | – | – | – | | |
| – | 842 | – | 840 | | | | 843 | 839 | | 838 | – | 840 |
| 845 | | – | 845 | | | | 848 | 845 | | | | |
| 850 | 851 | 851 | 851 | | | | 851 | | | | | 852 w |
| – | 855 | – | 854 | | | | – | | | | | |
| – | 863 | 864 | 865 | | | 865 | 862 | 861 | | | | |
| | | | | | | | | | | 870 | | |
| | 887 | 886 | 885 | | 889 | 887 | | | | 893 | 888 | |
| | | 895 | 895 | | 891 | | | | | | | 899 w |
| | | 904 | 904 | 905 s | | | | | | | 909 | |
| 914 | | 916 | 913 | | | | | | | | | 918 |
| | | | | | | | | | | | 933 | |
| 932 | | 934 | 936 | | 937 | 940 | | 935 | | | 937 | 937 |
| 954 w | | | 955 | | | | | 957 | 958 | | 957 | 950 |
| 972 | | 974 | 971, 975 | 975 | 974 w | | | | | | 965 | 964 |
| 979 | | | 979 | | | | | | | | | 980 |
| 989 w | | | | 986 s | | | | | | | 989 | |

Commissioned LIF data of MgTBP by Dr. Brumbaugh column (1) and Raman data (column 13) supplied by Drs. Markwort and N. Spingarn.

**Table 1e** Vibrational Frequencies of Tetrabenzoporphyrins plus Magnesium Porphyrin (MgP)

| MgTBP | | | | | | | CdTBP | ZnTBP | | H2TBP | MgP | MgTBP |
|---|---|---|---|---|---|---|---|---|---|---|---|---|
| | [7] | | Johnson | | | Spingarn | [7] | [74] | [13] | Johnson | [26] | |
| Bare | n-Octane 4 K | | Quasiline 77 K | | FTIR | Reflection 300 K | 4 K | Bare | Matrix | Matrix | 4 K | Raman |
| $S_1$ | $S_1$ | $S_0$ | $S_0$ | $S_1$ | $S_0$ | $S_0$ | $S_0$ | $S_1$, $S_2$ | (S) | (S) | (S) | $S_0$ |
| 1 | 2 | 3 | 4 | 5 | 6 | 7 | 8 | 9 | 10 | 11 | 12 | 13 |
| 998 | | 995 | 997 | | 1001 | | | | | | 994 | |
| | | | | | | | | | | | 998 | |
| | 1007 | 1006 | 1007 | 1006 | | | 1008 | | | | 1007 | |
| | | | 1011 | 1011 | | | | | | 1010 IR | 1011 | 1010 |
| 1019 | | | 1017, 1018 | | 1015 | 1015 | 1015, 1018 | | | | 1013 | 1018 |
| 1023 | | | | | | | | | | | | |
| 1026 | | 1026 | 1025 | | | | | | | | | |
| 1030 | | | 1029 | | | | | | | | | |
| | | | 1034 | | | | | 1032 | | 1033 | | |
| | 1037 | 1036 | 1037 | 1035, 1037 | 1036 | | | | | 1036 | | 1035 |
| | 1043 | | 1040 | | 1040 | | | | | | 1041 | 1043 |
| | | 1056 | | | 1057 | 1052 | | | | 1065 | 1057 | |
| | 1063 | 1060 | 1061 | | | | | | | | 1060 | 1059 |
| | 1067 | | | | | | 1070 | | 1066 | 1070 | | |
| | | | | | | | 1080 | | 1074 | | | |
| 1079 | 1078 | | 1080 | 1081 | | | | | | | | 1080 |
| | 1093 | | 1092 | | | | 1093 | | | 1092 | 1091 | 1095 |
| 1099 | 1098 | | 1097 | | | | 1096 | | | 1098, 1101 | | |
| | 1105 | 1106 | 1106 | 1105 | | | 1106 | | | | | |
| 1109 | | | 1110 | 1110 | | | 1110 | | | 1110 | | |
| 1111 | 1111 | 1111 | 1112 | 1112 | 1111 | 1111 | 1113 | 1114 | | 1112 | | |
| 1116 | 1116 | | 1116 | 1117 | 1119 | | 1117 | | | | 1116 | 1117, 1118 |
| | | | | | 1121 | | | 1121 | 1123 | | | |
| | 1127 | | 1128 | | | | | | | | | |
| 1131 | 1131 | | 1132 | 1133, 1136 | | | 1135 | | | | | |

Commissioned LIF data of MgTBP by Dr. Brumbaugh column (1) and Raman data (column 13) supplied by Drs. Markwort and N. Spingarn.

**Table 1f.** Vibrational Frequencies of Tetrabenzoporphyrins plus Magnesium Porphyrin (MgP)

| MgTBP | | | | | | | CdTBP | ZnTBP | | $H_2$TBP | MgP | MgTBP |
|---|---|---|---|---|---|---|---|---|---|---|---|---|
| | [7] | | Johnson | | | Spingarn | [7] | [74] | [13] | Johnson | [26] | |
| Bare | n-Octane 4 K | | Quasiline 77 K | | FTIR | Reflection 300 K | 4 K | Bare | Matrix | Matrix | 4 K | Raman |
| $S_1$ | $S_1$ | $S_0$ | $S_0$ | $S_1$ | $S_0$ | $S_0$ | $S_0$ | $S_1, S_2$ | *(S)* | *(S)* | *(S)* | $S_0$ |
| 1 | 2 | 3 | 4 | 5 | 6 | 7 | 8 | 9 | 10 | 11 | 12 | 13 |
| 1153 | | | 1153 | | | | 1150 | 1151 | | | | 1151 |
| | | | 1156 | | | | 1155 | | | 1155 | | 1154 |
| | 1158 | 1158 | 1158 | 1158 | 1157 | | | 1160 | 1159 | | 1159 | |
| | 1162 | | | 1164 | | | | | | | 1164 | |
| 1167 | | | | | | | 1167 | | | | | |
| | | 1170 | | 1169 | | | 1169 | | | | | |
| 1175 | 1173 | | 1174 | 1172 | | | 1173 | | | 1174 | | |
| | 1179 | 1184 | | 1184 | | | 1179 | 1181 | | 1183 | 1178 | 1179 |
| 1185 | 1186 | 1187 | 1187 | | | | | 1187 | | 1187 | 1189 | |
| | 1187 | | 1189 | | | | | | | | | |
| | 1190 w | | | | | 1191 | | | | | | |
| 1195 | 1192 w | | 1195 | 1195 | | | 1196 | | | | | 1196 |
| | 1200 | | 1200 | 1200 | 1202 | | | | | | | |
| 1212 | 1212 | 1214 | 1210 | | 1214 | | 1206 | 1208 | | 1209 | | |
| 1219 | 1219 | 1219 | 1218, 1220 | 1220 | 1217 | | | | | 1215 | 1219 | |
| 1221 | | | 1223 | | | | | | | | | |
| | | | 1226 | | | | | | | | | 1225, 1226 |
| | 1229 | | 1231 | 1231 | | | 1235 | | | 1234 | 1235 | |
| 1237 | 1238 | | 1237 | | 1236 | 1236 | | | 1238 | | | 1240 |
| | 1249 | 1249 | 1252 | 1251 | | | | | 1251 | | | |
| | 1260 | 1258 | 1257, 1260 | | | | 1256 | | | | 1262 | 1258 w |
| 1264 | | | | | | | | | | | | 1264 w |
| 1266 | 1267 | | 1267 w | 1267 | | | 1268 | | | | | |
| 1271 | 1273 | 1271 | 1272 | 1270 | 1271 w | | | | | 1271 | | |
| 1276 | 1275 | 1275 | 1277 | | | | 1274 | 1274 | | 1274 | | |
| | | | | | 1291 | | | 1280 | | | | 1291 w |
| 1297 | 1297 | 1299 | 1296 | 1298 | 1296 | 1294 | 1299 | 1293 | | 1299 | | |

Commissioned LIF data of MgTBP by Dr. Brumbaugh column (1) and Raman data (column 13) supplied by Drs. Markwort and N. Spingarn.

**Table 1g.** Vibrational Frequencies of Tetrabenzoporphyrins plus Magnesium Porphyrin (MgP)

| MgTBP | | | | | | | CdTBP | ZnTBP | | $H_2$TBP | MgP | MgTBP |
|---|---|---|---|---|---|---|---|---|---|---|---|---|
| | [7] | | Johnson | | | Spingarn | [7] | [74] | [13] | Johnson | [26] | |
| Bare | *n*-Octane 4 K | | Quasiline 77 K | | FTIR | Reflection 300 K | 4 K | Bare | Matrix | Matrix | 4 K | Raman |
| $S_1$ | $S_1$ | $S_0$ | $S_0$ | $S_1$ | $S_0$ | $S_0$ | $S_0$ | $S_1$, $S_2$ | (S) | (S) | (S) | $S_0$ |
| 1 | 2 | 3 | 4 | 5 | 6 | 7 | 8 | 9 | 10 | 11 | 12 | 13 |
| 1308 | 1309 | | 1309 | | | | 1310 | | | 1307 | 1305 | 1310 |
| 1313 | 1314 | 1315 | 1313, 1316 | | | | | | 1315 | | | |
| 1320 | | | 1321 | | 1321 | | 1322 | | | | 1319 | |
| 1323 | 1324 | | | | | | | 1323 | 1324 | | 1324 | 1323 |
| 1325 | 1327 | | 1327? | 1329 | 1329 | 1329 | | | | | | 1330 |
| | | | | | | | | | | 1331 | | |
| | | 1336 | 1336 | 1334 | | | 1335 | | 1334 | | | |
| 1337 | | 1338 | 1338 | 1339 | 1338 | | 1337 | | | 1337 | | |
| End of data | | | | | | | | | 1341 | | | |
| | 1342, 1345 | 1345 | 1342, 1347 | | 1348 w | | 1346 | | | 1346 | | |
| | 1353 | 1353 | 1352 | | 1356 w | | 1353 | 1350 | 1351 | 1356 | | |
| | | | | 1365 | 1362 | | | | | | 1363 | |
| | 1370 | 1368 | 1367, 1370 | | | | | | | | 1369 | 1367, 1370 |
| | 1374 | | 1374 | 1375 | 1373 | | 1373 | | | | | |
| | 1380 | 1383 | | | | | 1382 | | | 1380 | | 1378 |
| | 1387 | 1387 | 1388 | 1389 | 1387 | | 1391 | 1388 | | | | |
| | 1396 | | 1396, 1398 | 1394 | 1396 | | | | | 1397 | 1397 | 1394 |
| | 1402 | | 1400 | 1401 | | | | | | | | |
| | | 1417 | | | 1418 | 1413 | 1412 | | | | | 1409 w |
| | 1420 | | 1420 | 1420 | | | | | | 1420 | | |
| | 1426 | 1423 | | 1425 | 1427 w | | | | | 1426 | | |
| | 1435 | 1435 | 1434, 1436 | 1435, 1438 | 1439 | | | | | 1433 | 1443 | 1436 |
| | 1451 | 1452 | 1450 | 1451 s | 1446 | | | | | | 1457 | |
| | | | | 1456 | 1458 | 1454 | | | 1456 | | | |
| | | | 1465 | 1466 | 1466 | | | 1466 | | 1468 | | 1464 w |
| | | 1473 | 1473 | 1473 | 1474 | 1476 | | | | 1474 | | 1475 w |

Commissioned LIF data of MgTBP by Dr. Brumbaugh column (1) and Raman data (column 13) supplied by Drs. Markwort and N. Spingarn.

**Table 1h.** Vibrational Frequencies of Tetrabenzoporphyrins plus Magnesium Porphyrin (MgP)

| MgTBP | | | | | | | CdTBP | ZnTBP | | $H_2$TBP | MgP | MgTBP |
|---|---|---|---|---|---|---|---|---|---|---|---|---|
| | [7] | | Johnson | | | Spingarn | [7] | [74] | [13] | Johnson | [26] | |
| Bare | *n*-Octane 4 K | | Quasiline 77 K | | FTIR | Reflection (300 K) | 4 K | Bare | Matrix | Matrix | 4 K | Raman |
| $S_1$ | $S_1$ | $S_0$ | $S_0$ | $S_1$ | $S_0$ | $S_0$ | $S_0$ | $S_1$, $S_2$ | (S) | (S) | (S) | $S_0$ |
| 1 | 2 | 3 | 4 | 5 | 6 | 7 | 8 | 9 | 10 | 11 | 12 | 13 |
| | 1481 | | 1482 | | 1481 | | | | | | | |
| | | | | | 1489 | | | | | | 1487 | 1490 w |
| | | | | | 1496 | | 1495 | | | | 1496 | 1496 |
| | | 1502 | | 1502 | | | End of data | | | | 1500 | 1502 |
| | | | 1509 | 1508 | 1507 | | | | 1510 | 1502 | 1508 | 1508 |
| | | | 1518 | | 1518 | | | 1516 | | 1521 | | 1515 |
| | | 1536 | 1535 | 1534 s | 1533 | 1536 | | 1528 | | 1532 | | |
| | | | | 1540 | 1541 | | | End of data | | 1538, 1540 | | |
| | | 1551 | 1549 | | 1550 | | | | | 1550 | 1546 | 1549, 1555, 1552 |
| | | | | 1557 | 1559 | | | | | | 1563 | 1561 |
| | | 1566 | 1568 | | 1568 | | | | 1570 | | 1566, 1570 | 1568 |
| | | 1585 | 1588 | 1588 | 1585 | | | | | | | 1586 |
| | | | 1600 | 1598 | 1593 | | | | | 1601 | 1601 | 1598, 1596 |
| | | | | | | 1606 | | | | | 1609 | |
| | | 1615 | 1613 | | 1616 | | | | | 1617 | 1611 | |
| | | 1618 | 1618 | | | | | | | | 1619 | 1620 |
| | | End of data | 1626 | 1624 | 1626 | | | | 1626 | 1624, 1627 | | |
| | | | 1636 | 1634 | 1636 | | | | End of data | End of data | | |
| | | | 1642, 1644 | | 1645 | | | | | | 1647 | |
| | | | 1652 | | 1653 | | | | | | End of data | |
| | | | 1663, 1666 | 1660 | 1663 | | | | | | | 1669 |
| | | | 1674 | | 1676 | | | | | | | |
| | | | 1683 | | 1684 | | | | | | | 1687 |
| | | | 1698 | 1696 | 1697 | | | | | | | |

Commissioned LIF data of MgTBP by Dr. Brumbaugh column (1) and Raman data (column 13) supplied by Drs. Markwort and N. Spingarn.

**Table 1i**.  Vibrational Frequencies of Tetrabenzoporphyrins plus Magnesium Porphyrin (MgP)

| MgTBP | | | | | | | CdTBP | ZnTBP | | $H_2$TBP | MgP | MgTBP |
|---|---|---|---|---|---|---|---|---|---|---|---|---|
| | [7] | | Johnson | | | Spingarn | [7] | [74] | [13] | Johnson | [26] | |
| Bare | n-Octane 4 K | | Quasiline 77 K | | FTIR | Reflection 300 K | 4 K | Bare | Matrix | Matrix | 4 K | Raman |
| $S_1$ | $S_1$ | $S_0$ | $S_0$ | $S_1$ | $S_0$ | $S_0$ | $S_0$ | $S_1$, $S_2$ | (S) | (S) | (S) | $S_0$ |
| 1 | 2 | 3 | 4 | 5 | 6 | 7 | 8 | 9 | 10 | 11 | 12 | 13 |
| | | | 1719 | | 1717 | | | | | | | 1722 s |
| | | | 1728 | | | | | | | | | |
| | | | 1734 | 1734 s | 1734 | | | | | | | |
| | | | 1740 | | 1745 | | | | | | | 1739 w |
| | | | 1750 | | 1749 | | | | | | | 1750 |
| | | | 1760 | | 1762 | | | | | | | 1765 s |
| | | | 1769, 1775 | | 1772 | | | | | | | |
| | | | | | 1782 | | | | | | | 1779 |
| | | | | | 1792 | | | | | | | End of data |
| | | | 1801, 1805 | | 1802 | | | | | | | |
| | | | 1810 | | 1811 | | | | | | | |
| | | | 1816 | | 1828 | | | | | | | |
| | | | 1847 | | 1845 | | | | | | | |
| | | | | | 1859 | | | | | | | |
| | | | 1872 | | 1869 | | | | | | | |
| | | | | | 1946 | | | | | | | |
| | | | | | | | | | | | | |
| | | | | | | 3050 | | | | | | |

Commissioned LIF data of MgTBP by Dr. Brumbaugh column (1) and Raman data (column 13) supplied by Drs. Markwort and N. Spingarn.

**Table 2.** Lab Spectra vs. DIBs

| Probable DIBs | Lab Wide | Notes | n | Lab Å |
|---|---|---|---|---|
| 4131 | | | 6 | 4130.0 |
| 4169.9 | | | 5 | 4170.5 |
| 4175 | | H | 4 | 4175.0 |
| 4186 | | b | 5 | 4184.9 |
| 4187.5 | | b | 2 | 4187.7 |
| 4268 | | | 2 | 4268.0 |
| 4277 | | | 6 | 4276.5 |
| 4283 | | vs | 5 | 4282.9 |
| 4286 | | | 1 | 4287.5 |
| 4301 | | | 2 | 4301.5 |
| 4305 | | | 2 | 4304.5 |
| 4308 | | | 2 | 4307.8 |
| 4314 | | | 4 | 4315.1 |
| 4320 | | b | 3 | 4320.4 |
| 4326 | | | 6 | 4325.5 |
| 4329 | | | 3 | 4329.5 |
| 4333 | | | 1 | 4334.1 |
| 4350 | | | 1 | 4351 |
| 4357 | | | 1 | 4357.5 |
| 4426 | | | 1 | 4424.8 |
| **4428** | | **H**/Soret | **11** | **4428.2** |
| 4437.5 | | | 2 | 4437.1 |
| 4440 | | | 1 | 4438.8 |
| 4444 | | | 4 | 4444.5 |
| 4447.5 | | | 2 | 4447.2 |
| 4453 | | | 3 | 4454.1 |
| 4512.8 | | | 1 | 4512.3 |
| 4522.5 | | | 1 | 4521.6 |
| 4528 br | | | 3 | 4527 |
| 4550 | | | 1 | 4551.2 |
| 4581 br | | | 1 | 4579.8 |
| 4614 | | | 2 | 4613.9 |
| 4621 br | | | 1 | 4622.6 |
| 4631 | | | 4 | 4631.5 |
| 4635.8 | | | 1 | 4636.2 |
| 4642.8 | | | 1 | 4642.6 |
| 4682.5? | | | 1 | 4681.3 |
| 4710.5? | | | 1 | 4710.4 |
| 5264.8 | | | 2 | 5265.5 |
| 5711.5 | | | 1 | 5712.4 |
| | | | 3 | 5755.8 |

| Probable DIBs | Lab Wide | Notes | n | Lab Å |
|---|---|---|---|---|
| 5762.5 | | | 3 | 5761.8 |
| 5768 | | | 1 | 5767.5 |
| 5797vs | | H | 5 | 5798 |
| 5813.5 | | | 4 | 5813.8 |
| 5844 | | H | 1 | 5843.4 |
| | | | 1 | 5863.7 |
| 5872 | | | 3 | 5872.1 |
| 5899 | | | 8 | 5898 |
| 6016 | | | 1 | 6017.1 |
| 6030 | | | 4 | 6032.4 |
| 6038 | | | 3 | 6039 |
| 6071 | | | 1 | 6069.8 |
| 6078 | | | 2 | 6078 |
| 6247.5 | | | 1 | 6248 |
| | | | 1 | 6254 |
| | | | 1 | 6258.8 |
| | | | 1 | 6264 |
| 6270 | | H | 4 | 6269.4 |
| 6272 | | | 1 | 6273.5 |
| 6279 | | | 7 | 6278.2 |
| 6281 | | | 5 | 6282.0 |
| 6284 | | H | 4 | 6284.0 |
| 6288 | | | 4 | 6288.5 |
| 6290 | | absorption | 6 | 6291 |
| | | | 5 | 6294.1 |
| | | | 10 | 6297.3 |
| | | | 4 | 6301.6 |
| 6307 | | H | 2 | 6306.7 |
| 6314 | | H | 18 | 6314.6 |
| 6318 | | H | 6 | 6318.3 |
| | | | 7 | 6322.9 |
| | | | 10 | 6328.4 |
| 6331 | | | 1 | 6331.1 |
| 6335 | | | 1 | 6333 |
| 6345 | 6345 | | 6 | 6345.7 |
| 6354 | 6354 | H | 1 | |
| 6357?em | | | 7 | 6355.9 |
| 6362 | 6361 | H | 1 | |
| 6365 | 6365 | | 5 | 6364.7 |
| 6367 | | H | 4 | 6367.3 |
| | 6370 | | 2 | 6373.7 |
| 6376 | | H | 1 | 6377 |

| Probable DIBs | Lab Wide | Notes | n | Lab Å |
|---|---|---|---|---|
| 6379 | | H | 1 | 6379 |
| 6383? | | | 1 | 6381 |
| 6397 | | H | 1 | 6397 |
| 6400 | | | 1 | 6398.8 |
| 6410 | | | 1 | 6409.5 |
| 6432.8 | | | 1 | 6431.7 |
| 6439.5 | | H | 2 | 6441.2 |
| 6445.3 | | H | 3 | 6445.8 |
| 6449.3 | | H | 1 | 6451.0 |
| 6456 | | | 1 | 6455.5 |
| 6466.8? | | | 5 | 6465.7 |
| 6474 | | | 9 | 6476.9 |
| 6482 | | | 8 | 6481.6 |
| 6489 | | | 5 | 6488.7 |
| 6492 | | | 1 | 6491.5 |
| 6495 wide | | | 10 | s6494.9 |
| 6502 | | | 2 | 6502.8 |
| 6507 | 6508 | | 2 | 6508.8 |
| 6509 | | | 9 | 6511.5 |
| | | | 12 | 6513 |
| | 6515 | | 9 | 6514.5 |
| | | | 2 | 6530.5 |
| | | | 5 | 6533.0 |
| | 6538 | | 5 | 6536.7 |
| | | | 2 | 6552.4 |
| | | | 4 | 6562.6 |
| | | | 1 | 6568.7 |
| | | | 3 | 6574.0 |
| | | | 1 | 6576 |
| 6578.5 | | | 4 | s6578.2 |
| 6583 | | | 8 | 6582.0 |
| | | | 1 | 6587.0 |
| | | | 1 | 6590 |
| | | | 2 | 6591.5 |
| 6597 | | | 5 | 6595.5 |
| 6600 | | | 1 | 6600 |
| 6603 w | | | 8 | 6602.9 |
| | 6604 | | 2 | 6605.4 |
| 6609? | 6608 | | 2 | 6609.2 |
| 6611 | | | 2 | 6611.3 |
| 6614 | | | 2 | 6614.6 |
| | | | 2 | 6618.0 |

| Probable DIBs | Lab Wide | Notes | n | Lab Å |
|---|---|---|---|---|
| | | | 9 | 6620.7 |
| 6622.5 | 6621 | | 2 | 6623.5 |
| 6627 | 6627 | | 4 | 6627.9 |
| 6633 | | | 1 | 6632.6 |
| 6635.3 | | | 3 | 6635.1 |
| 6638 w | | | 2 | 6638.5 |
| 6642 | | | 1 | 6640.0 |
| 6649 | wide | | 4 | 6647.6 |
| | | | 2 | 6650.6 |
| 6654 | | | 3 | 6654.4 |
| 6657 | | | 3 | 6656.3 |
| | | | 4 | 6658.4 |
| 6660.6 | | H | 1 | 6662.6 |
| 6672 | | | 2 | 6672.7 |
| | | | 1 | 6677 |
| | | | 1 | 6680.2 |
| 6689.4 | | | 3 | 6690.2 |
| 6694.5 | | | 4 | 6694.5 |
| 6704.2 | | | 1 | 6705 |
| 6740.8 | | H | 1 | 6740 |
| 6743.5 | | | 1 | 6744 |
| 6746 | | | 2 | s6747.3 |
| 6750.5 | | | 1 | 6749 |
| 6752.7 | | | 1 | 6752.6 |
| 6756 | | | 2 | 6755.3 |
| 6758 | | | 1 | 6758 |
| 6775 | | | 1 | 6774.9 |
| 6792.4 | | H | 1 | 6790.7 |
| 6795.2 | | H | 1 | 6794.6 |
| 6809.3 | | H | 3 | 6808.6 |
| 6811.1 | wide | H | 3 | 6811.0 |
| 6823.3 | | H | 1 | 6824.2 |
| 6832.8 | | H | 1 | 6832.2 |
| 6847.6 | | H | 4 | s6848.3 |
| 6852.5 | | H | 3 | s6853.3 |
| | | | 1 | 6865.1 |
| | | | 1 | 6882.2 |
| | | | 3 | 6889.5 |
| | | | 3 | s6892.6 |
| | 6906 | | 2 | 6908.2 |
| 6944.5 | 6931 | H | 2 | 6944.6 |
| | | | 1 | 6970.9 |

| Probable DIBs | Lab Wide | Notes | n | Lab Å |
|---|---|---|---|---|
| 6993 | | H | 2 | 6993.1 |
| 7023.3 | | | 1 | 7022.9 |
| 7027.4 | | | 2 | 7027.5 |
| 7037 | wide | | 1 | 7038 |
| 7050.2 | wide | | 1 | 7050.2 |
| He | | | 3 | 7065.8 |
| v7068.2 | | | 2 | 7068.0 |
| 7069.5 | | | 1 | 7070.3 |
| 7075 | | | 1 | 7075.4 |
| 7088 | | | 1 | 7088 |
| 7282.3 | | | 1 | 7283.4 |

n= number of measurements
H= Herbig designated DIBs
br= broad
v.s.= very sharp
s= strong
w= weak

Column 1 lists probable and definitive DIBs taken from the published literature. Source for Herbig data, designated by “H” is found in reference [2]. The rest is from a survey paper, see [37]. Columns 2 and 5 list the totality of all the author’s quasiline Shpolskii lab data. Column 4 list the number of individual measurements taken for each of the data points. The author’s data in column 5 is the average value of the total number (n) of the measurements for each spectral data point.

**Table 3** Author's Shpolskii lab spectra of MgTBP at 77K in the blue spectral region (including the Soret band) and assigned vibronic transitions. Note: confirmation of the $S_2$ splitting.

| Notes | n | Å | m.e Å | 1/ λ (cm-1) | Origins 22444 $S_{2X}$ (cm-1) | Origins 22479 $S_{2Y}$ (cm-1) | Notes |
|---|---|---|---|---|---|---|---|
| | 6 | 4130.O | 0.3 | 24213 | 1769 | 1734 | |
| | 5 | 4170.5 | 0.4 | 23978 | 1534 | | |
| | 4 | 4175.O | 0.4 | 23952 | 1508 | 1473 | |
| b | 5 | 4184.9 | 0.5 | 23895 | 1451 | 1416 | |
| b | 2 | 4187.7 | 1 | 23879 | 1435 | 1400 | |
| | 2 | 4268.O | 0.7 | 23430 | 986 | | |
| | 6 | 4276.5 | 0.3 | 23384 | | 905* | ↑35 cm-1 |
| vs | 5 | 4282.9 | 0.3 | 23349 | 905* | | ↓ |
| | 1 | 4287.5 | 1 | 23324 | | 845 | |
| | 2 | 4301.5 | 0.9 | 23248 | | 769* | ↑ |
| | 2 | 4304.5 | 0.3 | 23231 | 787 | 752 | 34cm-1 |
| | 2 | 4307.8 | 1 | 23214 | 770* | 735 | ↓ |
| | 4 | 4315.1 | 0.6 | 23174 | | 695 | |
| b | 3 | 4320.4 | 0.6 | 23146 | 702 | | |
| | 6 | 4325.5 | 0.5 | 23119 | 675 | 640 | |
| | 3 | 4329.5 | 0.9 | 23097 | | 618 | |
| | 1 | 4334.1 | 1 | 23073 | 629 | 594 | |
| | 1 | 4351 | 1 | 22983 | 539 | 504* | ↑34 cm-1 |
| | 1 | 4357.5 | 1 | 22949 | 505* | | ↓ |
| | 1 | 4424.8 | 1 | 22600 | 156 | | |
| **Soret** | 11 | **4428.2** | 0.4 | 22583 | <139> | | |
| | 2 | 4437.1 | 0.3 | 22537 | 93 | | |
| | 1 | 4438.8 | 1 | 22529 | 85 | | |
| | 4 | 4444.5 | 0.3 | 22500 | | 21 | |
| | 2 | 4447.2 | 1 | 22486 | 42 | | |
| | 3 | 4454.1 | 1 | 22451 | | 28 | |

n= number of measurements
b= broad
vs= very sharp

**Table 4a** Author's Shpolskii lab spectra of MgTBP at 77K and its interpretation involving sites and tentative assignments of vibrational frequencies associated with the vibronic transitions.

| | | MgTBP Fluorescence at 77K | | Site I | | Site II | | Site III | | Site IV | | Site V | | MgTBP Absorption lines At 77K | |
|---|---|---|---|---|---|---|---|---|---|---|---|---|---|---|---|
| DIBs | | Narrow Lines | | **15913** | **15949** | **15882** | **15918** | **15944** | **15980** | **15872** | **15908** | **15810** | **15846** | | |
| | | | | **x** | **y** | **x** | **y** | **x** | **y** | **x** | **y** | **x** | **y** | | |
| | **n** | **A** | $cm^{-1}$ | $cm^{-1}$ | $cm^{-1}$ | $cm^{-1}$ | $cm^{-1}$ | $cm^{-1}$ | $cm^{-1}$ | $cm^{-1}$ | $cm^{-1}$ | $cm^{-1}$ | $cm^{-1}$ | **n** | |
| | | 6248 | 16005 | **-92** | | -123 | -87 | **-61** | -25 | -133 | -97 | **-195** | -159 | | |
| | | 6254 | 15990 | **-77** | **-41** | -108 | **-72** | | **-10** | **-118** | -82 | -180 | -144 | | |
| | 1 | 6258.8 | 15977 | | | | | | **0-0** | | | | | | |
| | | 6264 | 15964 | | | | | **-20** | | **-92** | | **-154** | **-118** | | |
| * | 4 | 6269.4 | 15950 | | **0 - 0** | | | | | **-78** | -42 | | | 0 | (6270.4) inferred |
| | 1 | 6273.5 | 15940 | -27 | **9** | | | **0-0** | **40** | | -32 | -130 | **-94** | 2 | 6273.2 |
| | 7 | 6278.2 | 15928 | | **21** | | | | | | -20 | **-118** | -82 | 10 | 6277.4 mean |
| | 5 | 6282.O | 15918 | | | | **0-0#** | 26 | 62 | | **-10** | -108 | -72 | 5 | 6282.1 |
| * | 4 | 6284.O | 15913 | **0-0** | | | | | | | | | | 5 | 6284.4 |
| | 4 | 6288.5 | 15902 | **11** | | | | **42** | **78** | | | | | | |
| | | | 15896 | | | | | | 84 | -24 | | **-86** | | 6 | 6291 |
| | 5 | 6294.1 | 15888 | **25** | **61** | | | | **92** | | **20** | **-78** | **-42** | 5 | 6295.1 |
| | 10 | 6297.3 | 15879 | 34 | | **0-0** | 39 | | 101 | | 29 | | **-33** | 8 | 6298.9 |
| | 4 | 6301.6 | 15869 | 44 | | | | | 111 | **0-0** | 39 | | | | |
| | | | 15864 | | 85 | 18 | | | 116 | | 44 | | **-18** | 6 | 6303.5 |
| * | 2 | 6306.7 | 15856 | 57 | **93** | **26** | **62#** | | | | | | -10 | | |
| | | | 15848 | | | | | | | | | | **0-0** | 4 | 6309.8 |
| * | 18 | 6314.6 | 15836 | **76.7** | **112.7** | **45.7** | 81.7 | 107.7 | 143.7 | 35.7 | **71.7** | -26.3 | **9.7** | | **End of Absorption Data** |

# Correlates with Ref. [11]

**Table 4b** Author's Shpolskii lab spectra of MgTBP at 77K and its interpretation involving sites and tentative assignments of vibrational frequencies associated with the vibronic transitions.

| DIBs | 77K MgTBP Quasiline Spectra Fluorescent Lines | | | | | Site I | | Site II | | Site III | | Site IV | | Site V | | Sevchenko* Fluorescence Quasilines [11] |
|---|---|---|---|---|---|---|---|---|---|---|---|---|---|---|---|---|
| | | | | | | 15913 | 15949 | 15882 | 15918 | 15944 | 15980 | 15872 | 15908 | 15810 | 15846 | |
| | | Wide | Narrow | | | x | y | x | y | x | y | x | y | x | y | |
| | n | Å | Å | m.e. | $cm^{-1}$ | $cm^{-1}$ | $cm^{-1}$ | $cm^{-1}$ | $cm^{-1}$ | $cm^{-1}$ | $cm^{-1}$ | $cm^{-1}$ | $cm^{-1}$ | $cm^{-1}$ | $cm^{-1}$ | |
| * | 6 | | 6318.3 | 0.3 | 15827 | **86** | | **55** | **91** | **117** | | **45** | | | | |
| | 3 | | 6322.2 | | 15817 | | | | | | | | | | | |
| | 7 | | 6322.9 | 0.8 | 15815 | **98** | **134** | | | **129** | | | **93** | | | |
| | **1** | | **6324.7** | | **15811** | | | **71** | | | | **61** | **97** | **0-0** | | |
| | **3** | | **6327.3** | | **15805** | **108** | **144** | **77** | **113** | | **175** | | | | **41** | |
| | 10 | | 6328.4 | 0.6 | 15802 | **111** | **147** | | **116** | | | | | | **44** | |
| | **3** | | **6331.1** | | **15795** | **118** | **154** | **87** | | **149** | **185** | **77** | **113** | | | |
| | | | 6333 | | 15790 | | **159** | **92** | | **154** | **190** | | **118** | **20** | | |
| | 6 | 6345 | 6345.7 | 0.5 | 15759 | **154** | **190** | | **159** | **185** | **221** | **113** | **149** | | **87** | |
| * | 1 | 6354 | | | 15738 | **175** | **211** | **144** | **180** | | **242** | | | **72** | | |
| * | 7 | | 6355.9 | 0.6 | 15733 | **180** | **216** | **149** | **185** | | **247** | | **175** | **77** | **113** | |
| * | 1 | 6361 | | | 15721 | | | **161** | **197** | | **259** | **151** | **187** | | **125** | |
| | 5 | 6365 | 6364.7 | 0.8 | 15711 | | **238** | **171** | | | | **161** | **197** | **99** | | |
| * | 4 | | 6367.3 | 0.6 | 15705 | **208** | **244** | | | **239** | **275** | **167** | | | | |
| | | 6370 | | | 15699 | | **250** | | | **245** | **281** | **173** | **209** | **111** | **147** | |
| | 2 | | 6373.7 | | 15689 | | **260** | | **229** | **255** | **291** | | **219** | | **157** | |
| * | 1 | | 6377 | | 15681 | **232** | | | **237** | | | | | | | **6377** |
| * | 1 | | 6379 | | 15676 | **237** | | | **242** | | **304** | **196** | **232** | | | |
| | | | 6381 | | 15672 | **241** | | **210** | **246** | | | | **236** | | **174** | |
| * | 1 | | 6397 | | 15632 | **281** | | | **286** | **312** | | **240** | **276** | | | |
| | 1 | | 6398.8 | | 15630 | **283** | | | **288** | | | **242** | | **180** | **216** | |
| | | | | | 15615 | **298** | **334** | | **303** | | | **257** | | **195** | **231** | **6404** |

**Table 4c.** Author's Shpolskii lab spectra of MgTBP at 77K and its interpretation involving sites and tentative assignments of vibrational frequencies associated with the vibronic transitions.

| DIBs | 77K MgTBP Quasiline Spectra Fluorescent Lines | | | | | Site I | | Site II | | Site III | | Site IV | | Site V | | Sevchenko* Fluorescence Quasilines [11] |
|---|---|---|---|---|---|---|---|---|---|---|---|---|---|---|---|---|
| | | | | | | 15913 | 15949 | 15882 | 15918 | 15944 | 15980 | 15872 | 15908 | 15810 | 15846 | |
| | | Wide | Narrow | | | x | y | x | y | x | y | x | y | x | y | |
| | n | Å | Å | m.e. | $cm^{-1}$ | $cm^{-1}$ | $cm^{-1}$ | $cm^{-1}$ | $cm^{-1}$ | $cm^{-1}$ | $cm^{-1}$ | $cm^{-1}$ | $cm^{-1}$ | $cm^{-1}$ | $cm^{-1}$ | |
| * | 1 | | 6790.7 | | 14726 | **1187** | **1223** | **1156** | | **1218** | | | | | | |
| * | 1 | | 6794.6 | | 14718 | **1195** | **1231** | **1164** | **1200** | **1226** | | **1154** | **1190** | **1092** | | |
| * | 3 | | 6808.6 | 0.6 | 14687 | **1226** | | **1195** | **1231** | **1257** | | **1185** | **1221** | | **1159** | |
| * | 3 | wide | 6811.O | 0.5 | 14682 | **1231** | **1267** | **1200** | **1236** | | **1298** | **1190** | **1226** | **1128** | **1164** | |
| * | | | | | 14671 | | **1278** | **1211** | | **1273** | **1309** | **1201** | **1237** | | **1175** | 6816 |
| * | 1 | | 6824.2 | | 14653 | **1260** | **1296** | **1229** | | | **1327** | **1219** | | **1157** | **1193** | **End of Data** |
| * | 1 | | 6832.2 | | 14636 | **1277** | **1313** | **1246** | | **1308** | | **1236** | **1272** | **1174** | **1210** | |
| * | 4 | | s6848.3 | 0.9 | 14602 | | **1347** | **1280** | **1316** | | | **1270** | | | | |
| * | 3 | | s6853.3 | 0.5 | 14592 | **1321** | | **1290** | **1326** | **1352** | | | **1316** | **1218** | | |
| | 1 | | 6865.1 | | 14566 | **1347** | **1383** | **1316** | **1352** | | | | | | | |
| | 1 | | 6882.2 | | 14530 | **1383** | **1419** | **1352** | **1388** | | **1450** | **1342** | | | **1316** | |
| | 3 | | 6889.5 | 0.6 | 14515 | **1398** | **1434** | **1367** | | | **1465** | | | **1295** | | |
| | 3 | | s6892.6 | | 14508 | | | **1374** | | **1436** | **1472** | | **1400** | | | |
| | 2 | 6906 | 6908.2 | 0.2 | 14476 | **1437** | **1473** | **1406** | | | | **1396** | | **1334** | **1370** | |
| | 2 | 6931 | 6944.6 | 1 | 14400 | | **1549** | **1482** | **1518** | | | **1472** | | | | |
| | 1 | | 6970.9 | | 14345 | **1568** | | **1537** | | **1599** | **1635** | | | **1465** | **1501** | |
| * | 2 | | 6993.1 | 2 | 14300 | **1613** | | **1582** | **1618** | **1644** | | | | | | |

**Table 5.** Reported and probable DIBs in the blue spectral region. Columns 4 and 5 list vibrational frequencies for each of the corresponding vibronic transitions. Vibrational frequencies are consistent with the frequencies listed in Table 1, and hence confirm the MgTBP designation.

| **Reference** | **DIBs(Å)** | **DIBs** | **22,444** | **22,479** | **FWHM** | **EW** | **New DIB ?** |
|---|---|---|---|---|---|---|---|
| | | $cm^{-1}$ | $S_{2x}$ $cm^{-1}$ | $S_{2y}$ $cm^{-1}$ | (Å) | | Y=yes |
| S | 4381 | 22,826 | 382 | ---- | | | Y |
| S | 4394 | 22,758 | --- | 278 | | | Y |
| S | 4401 | 22,722 | 278 | --- | | | Y |
| S | 4407.5 | 22,688 | 244 | 209 | | | Y |
| S | 4413 | 22,660 | 216 | 181 | | | Y |
| S | 4417 | 22,640 | 196 | 161 | | | Y |
| S | 4420 | 22,624 | 180 | 145 | | | ? |
| M em. | 4485 | 22,296 | 148 | 183 | | | |
| H | 4501.7 | 22,214 | 230 | 265 | 2.8 | 190 | |
| M em. | 4503 | 22,207 | 237 | --- | | | |
| M | 4525.5 | 22,097 | --- | 382 | | | Y |
| M | 4535 | 22,051 | 393 | --- | | | Y |
| J & D | 4595 | 21,762 | 751, 615 | 787, 646 | 28 | | |
| J & D | 4665.5 | 21,434 | 1010 | --- | 4 | | |
| H | 4726.4 | 21,158 | --- | 1321 | 3.4 | 175 | |
| J & D | 4727 | 21,155 | --- | 1324 | 4 | | |
| H | 4760 | 21,008 | 1481,1451 | 1518,1420 | 25 | 700 | |
| | | | 1420,1380 | | | | |
| J & D | 4761.7 | 21,001 | --- | 1478 | | | |
| H | 4762.7 | 20,996 | --- | 1483 | 3.1 | 160 | |
| | | | | | | | |

**Key**: S= T.P. Snow, et. al. 2002, Astrophys. J. 578 (2002) 877 [76]
H= G.H. Herbig, Annu. Rev. Astrophys. 33 (1995) 19 [2]
M= N.I. Morrell, N.R. Walborn, E.L. Fitzpatrick, Astro. Soc. of the Pac. 103 (1991) 341-346 [72]
J&D= P. Jenniskens, and F-X. Desert, Astron. Astrophys. Suppl. 106 (1994) 39 [37]

**Table 6.** DIBs and their corresponding vibronic transitions for HD29647. This is the interpretation of astronomical data based on the work of [7] and vibrational frequencies from Table 1.

| P & C [7] ( Å) | DIBs ( Å) | DIBs (cm$^{-1}$) | 15,962 origin | 15,992 origin |
|---|---|---|---|---|
| 5779 | 5780.7 | 17,299 | 1,337 | 1308 |
| | 5782.1 | 17,295 | 1,333 | |
| | 5784.2 | 17,288.5 | 1326.5 | 1296 |
| | 5785 | 17,286.1 | 1324 | 1294 |
| | 5791.2 | 17,267.6 | | 1275 |
| | 5792.4 | 17,264.0 | | 1272 |
| 5794 | 5794 em | 17,259 | 1297 | 1267 |
| | 5794.3 em | 17,258 | 1296 | 1266 |
| | 5795.1 em | 17,256 | 1294 | 1264 |
| | 5795.4 | 17255 | 1293 | 1263 |

**Table 7.** Abbreviated extension of Table 4, listing potential sites and the corresponding frequencies (from Tables 1e & 1f) associated with the indicated DIBs.

←------------------------------- **SITES** ---------------------------------------→

| DIBs | cm$^{-1}$ | 15,913 | 15,949 | 15,882 | 15,918 | 15,944 | 15,980 | 15,872 | 15,908 | 15,810 | 15,846 |
|---|---|---|---|---|---|---|---|---|---|---|---|
| 5797.11 | 17,250 | 1338 | ___ | 1368 | 1333 | ___ | 1271 | 1375 | 1342 | 1437 | 1400 |
| 5849.78 | 17,095 | 1184 | ___ | 1212 | ___ | 1153 | 1116 | 1223 | 1187 | ___ | 1249 |

**Table 8.** The blue fluorescence spectra measured from Fig. 1 of HD44179 (Schmidt, et al. [93]) and its interpretation. The electronic origin is at 25344.7 cm$^{-1}$. Column three lists the assigned vibrational frequencies corresponding to each of the vibronic transitions.

| RR Emission (Å) | RR Emission (cm$^{-1}$) | $\Delta\upsilon$ (cm$^{-1}$) |
|---|---|---|
| 3713 | 26,932 | 1587.7 |
| 3752 | 26,652 | 1308 |
| 3779 | 26,462 | 1117 |
| 3800 | 26,316 | 972 |
| 3851.5 | 25,964 | 619 |
| 3884 | 25,746.6 | 402 |
| 3908 | 25,588 | 243 |
| 3945.6 | 25,344.7 | 0-0 |
| 3979 | 25,132 | 213 |
| 4009 | 24,944 | 401 |
| 4044 | 24,728 | 617 |
| 4077 | 24,528 | 817 |
| 4119 | 24,278 | 1067 |
| 4136 | 24,178 | 1166 |
| 4167 | 23,998 | 1347 |
| 4255 | 23,499 | 1846 |

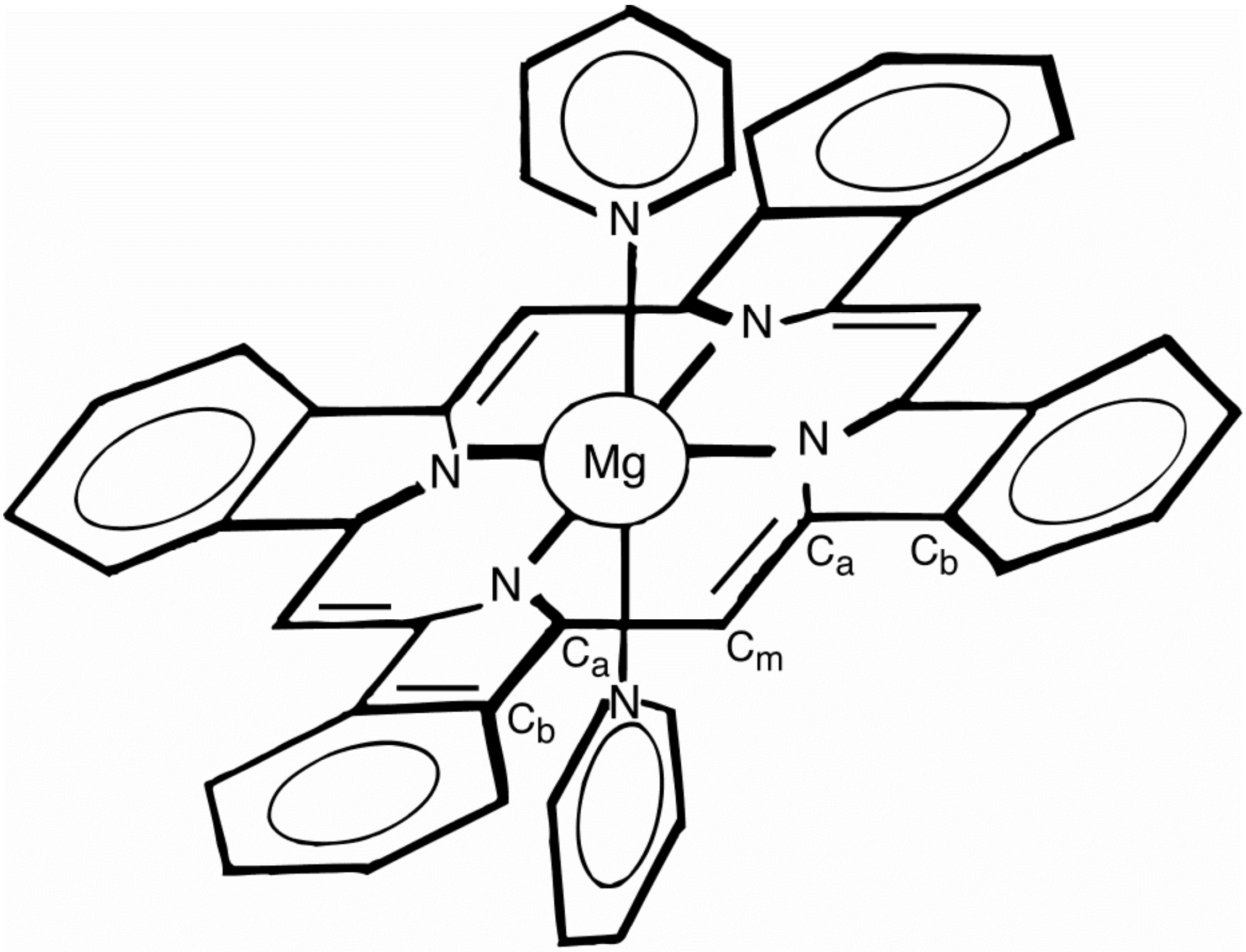

Figure 1. Dipyridyl magnesium tetrabenzoporphyrin ($MgC_{46}H_{30}N_6$)

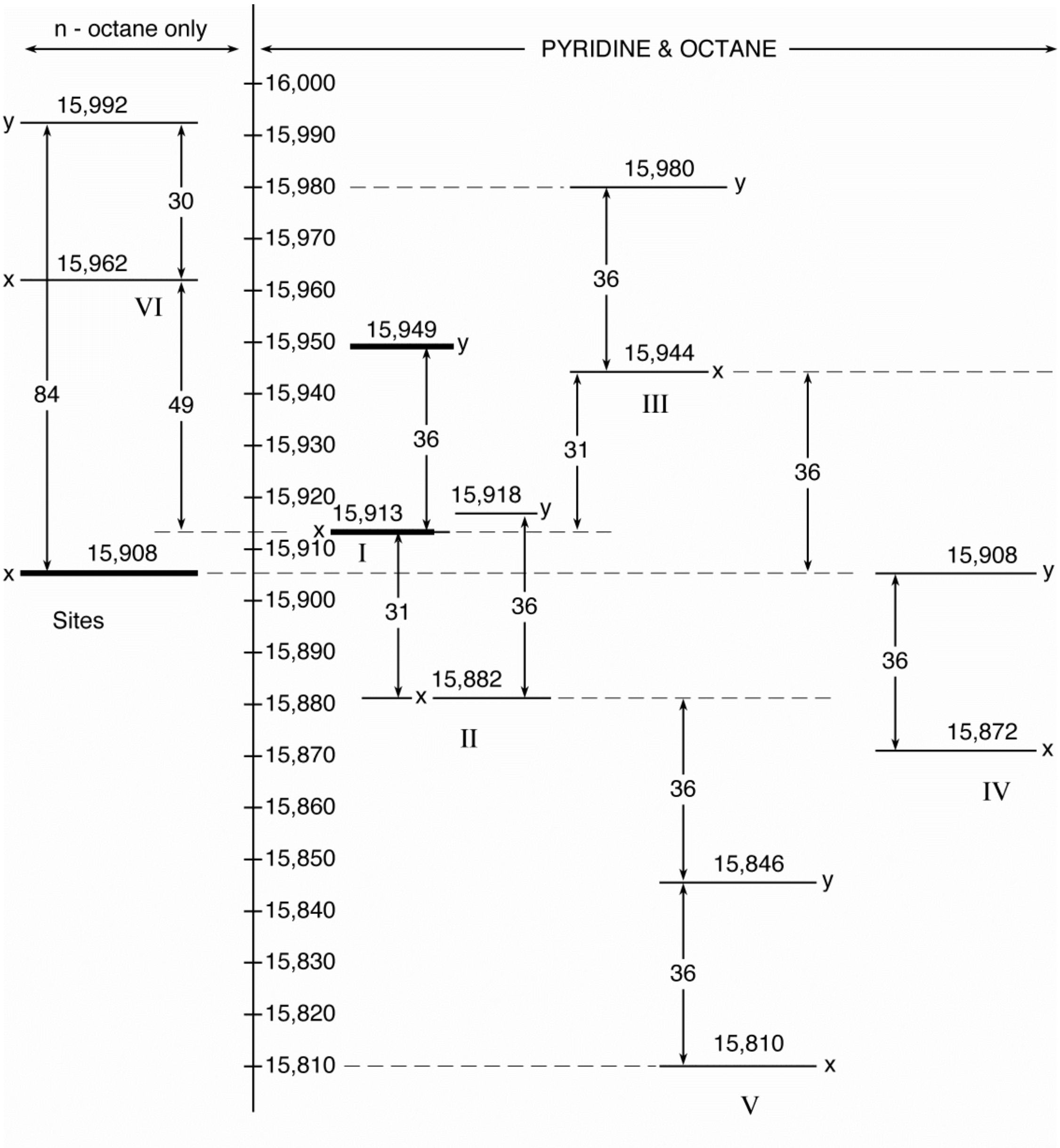

Figure 2 Energy (in $cm^{-1}$) level diagram (to scale) of "sites" of MgTBP in matrices - in n - octane and octane / pyridine mixtures for the $S_1$ electronic state (bare $S_1$ = 16,353 $cm^{-1}$)

Sites I thru V are derived from author's data. Site VI is derived from [7].

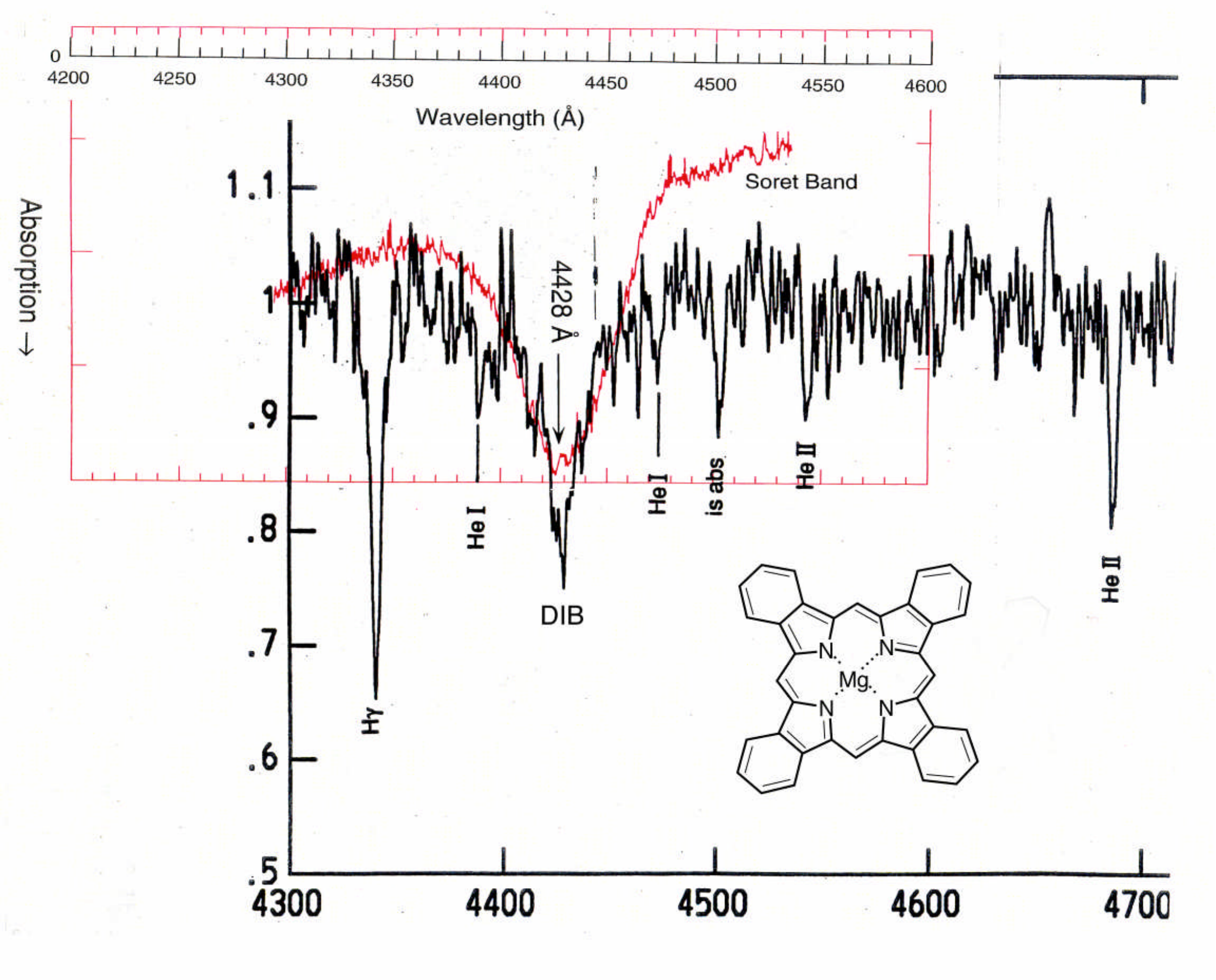

Figure 3. The superposition of MgTBP lab Soret band and the 4428Å DIB. Astronomical data from reference [72].
Superpositioned (red) lab Soret band is from author's data.

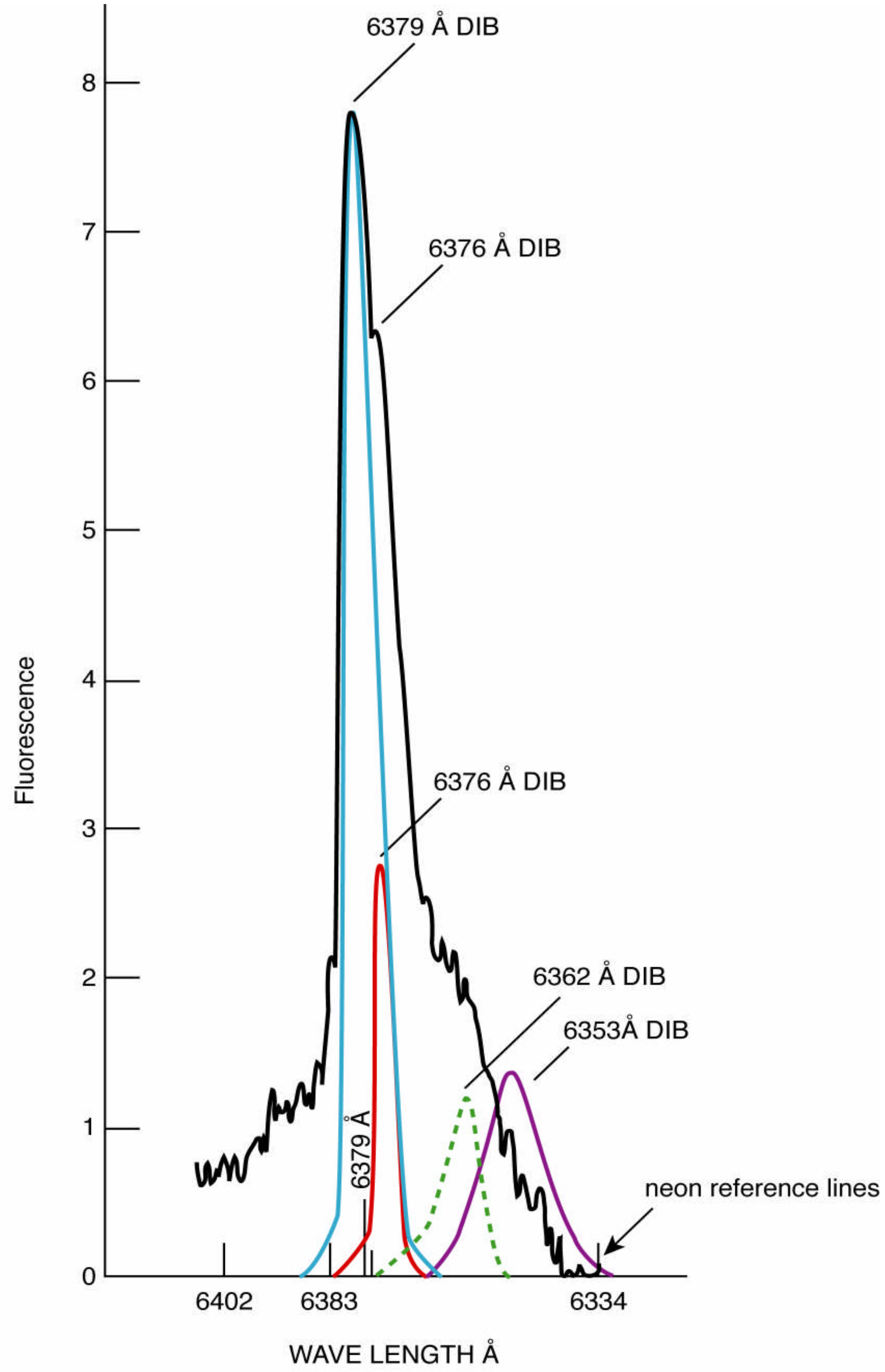

Figure 4. MgTBP in Shpolskii matrix at 77K, showing a lab band at 6379Å, which encompasses DIBs λλ 6379, 6376, 6362 and 6353

Author's lab data superimposed with astronomical data.

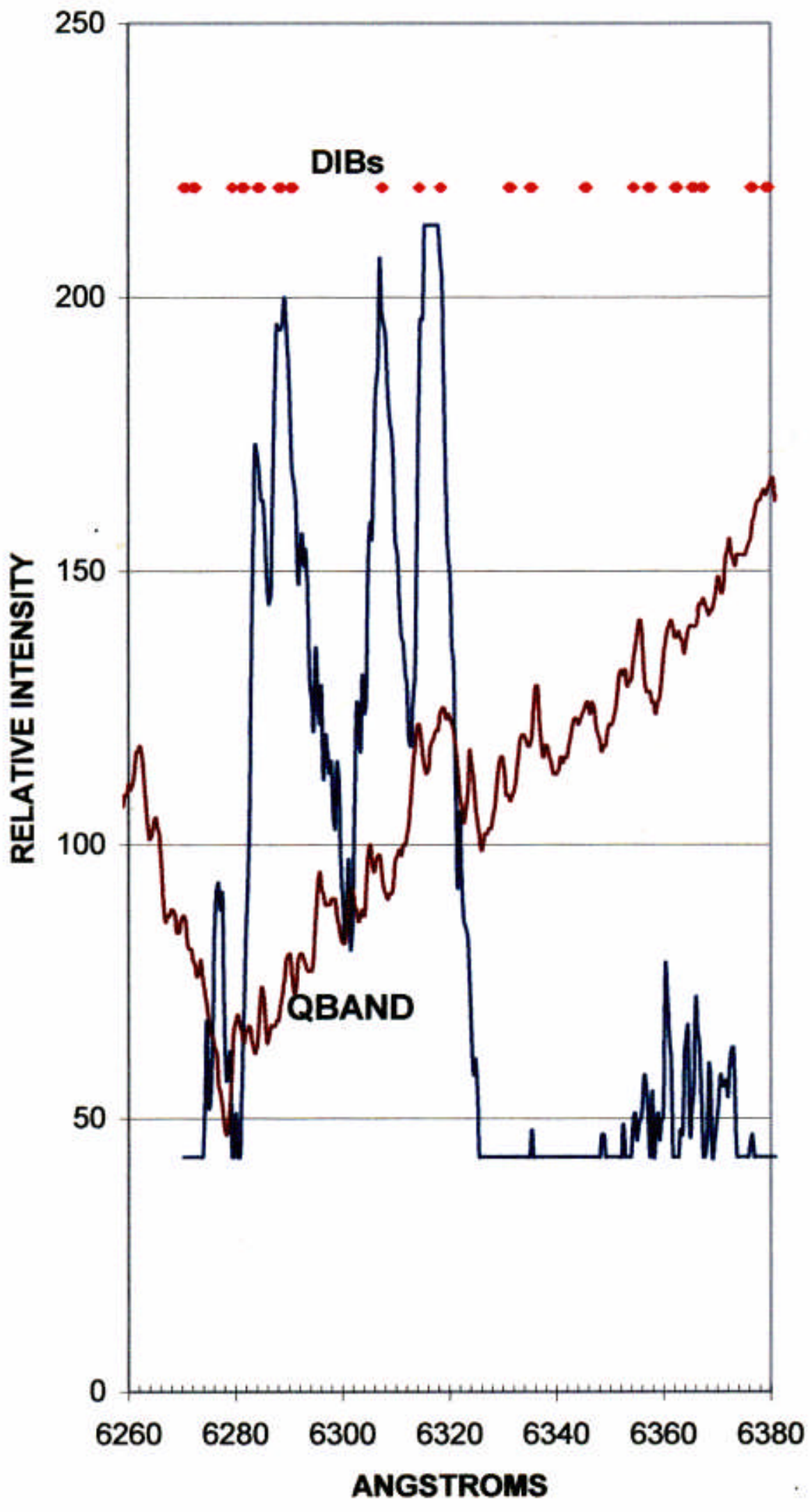

Figure 5. The superposition of two spectral lab bands (using rescanned data by David Weinkle). Note: the Q-band in absorption as well as an emission spectrum. The Q-band illustrates the superposition of an inhomogeneously broadened Q-band, plus a narrow Shpolskii band in the 6284Å region.

Spectral plates from author's data set.

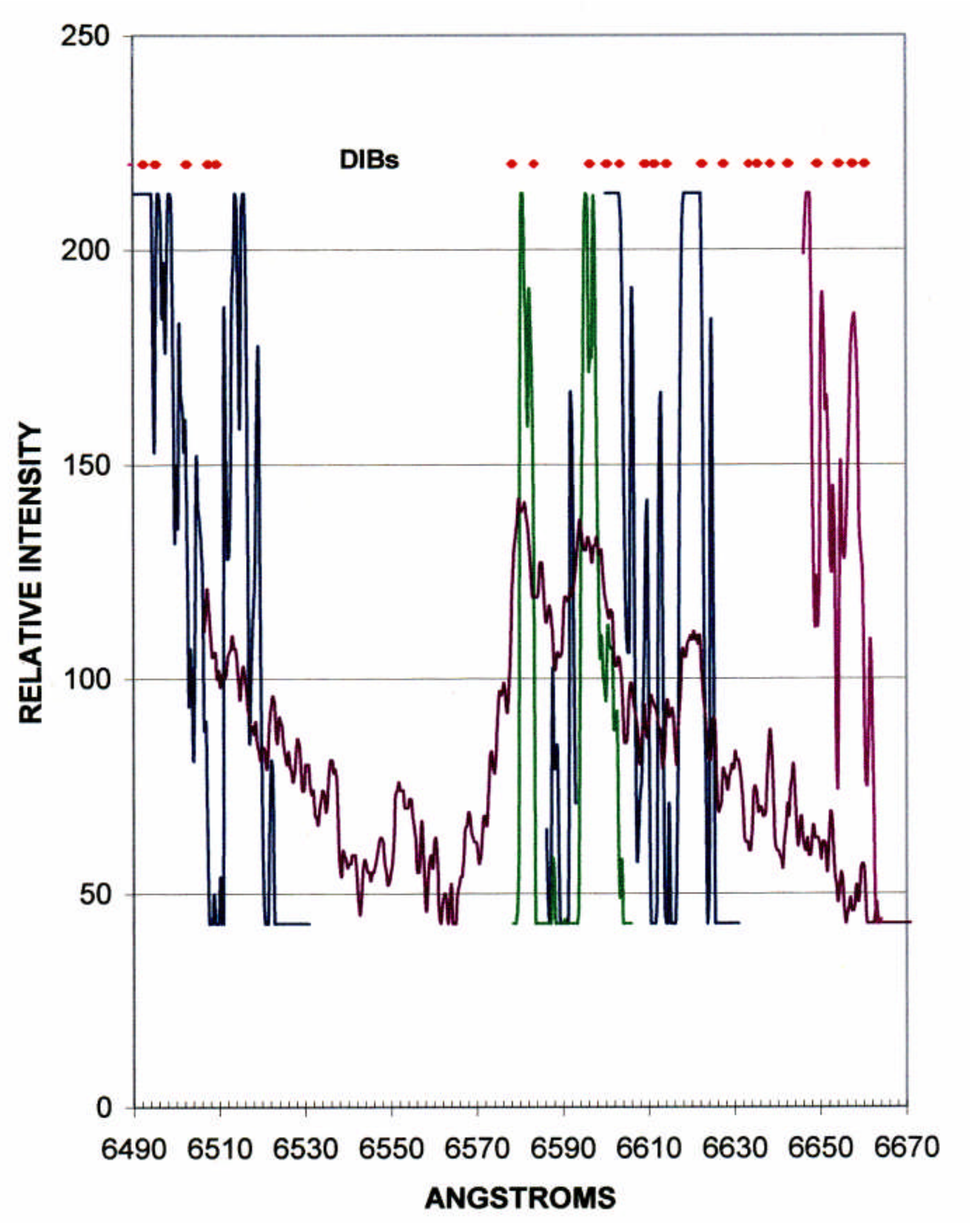

Figure 6. Six individual, superimposed Shpolskii spectral scans.
Note: the overlap of some of the individual spectral lines.
Also shown are the relevant, corresponding DIBs.
Spectral Data from author's data set.

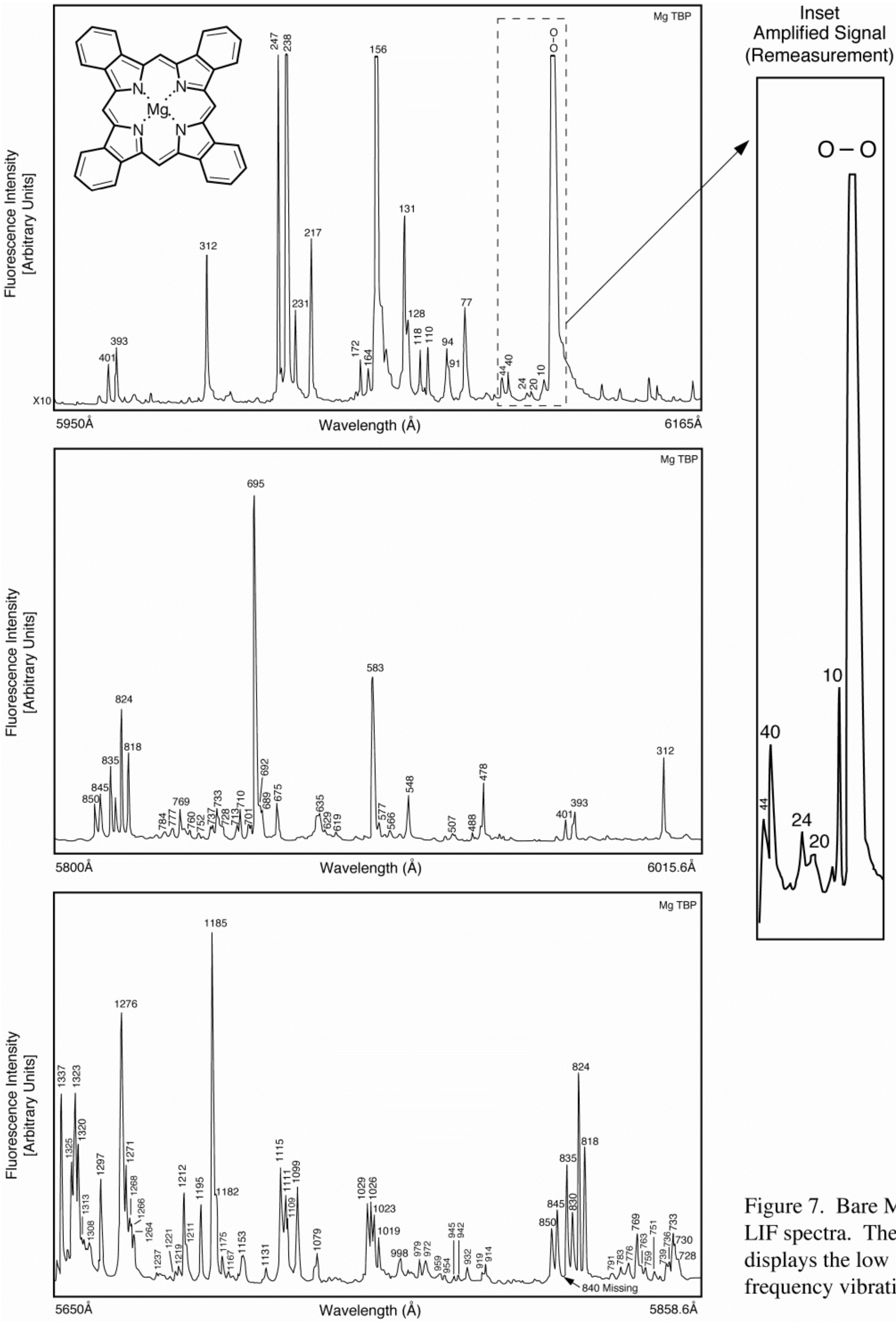

Figure 7. Bare MgTBP LIF spectra. The inset displays the low frequency vibrations.

This commissioned data was supplied by Dr. Don Brumbaugh

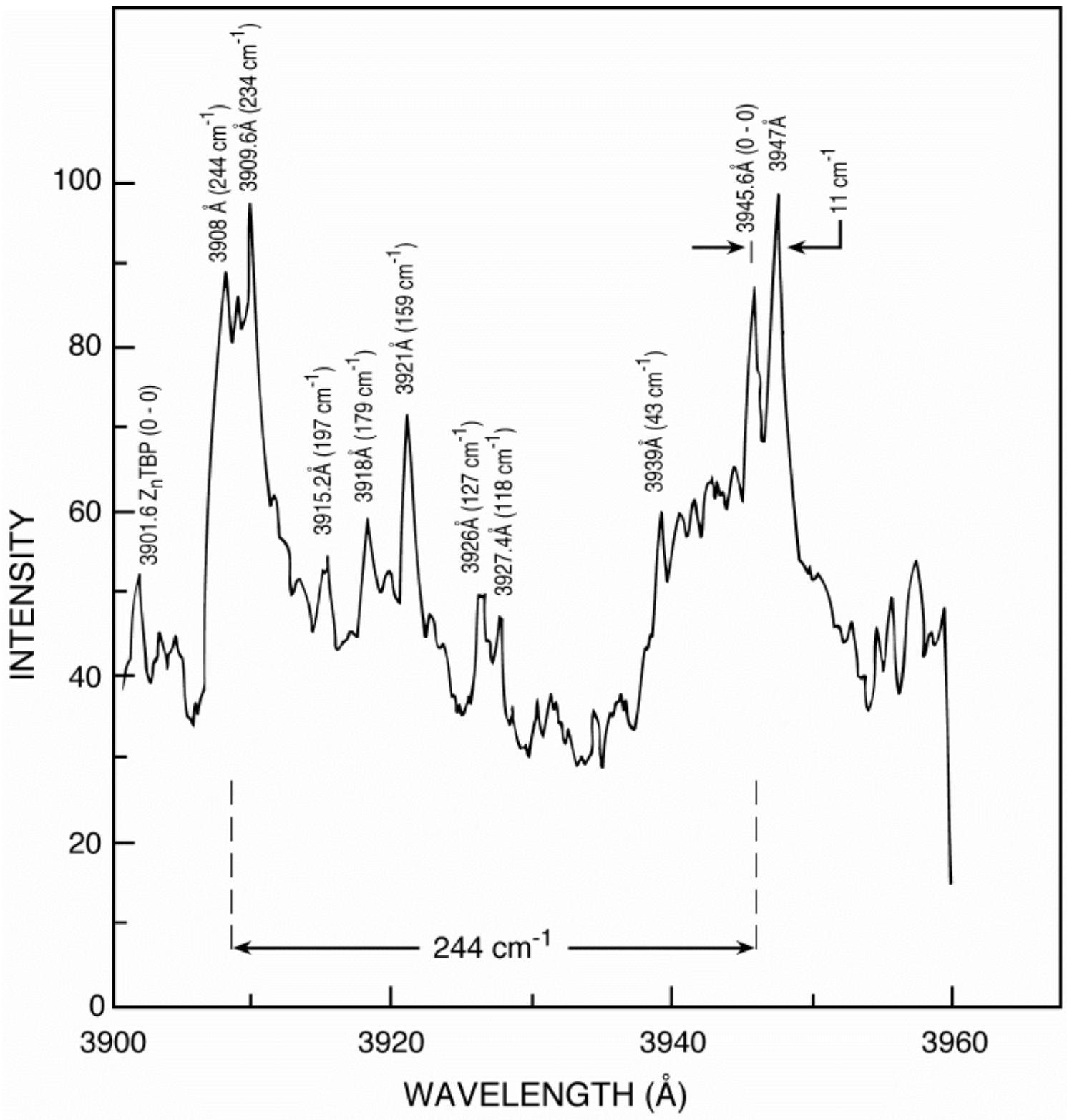

Figure 8a

The $S_2 \leftarrow S_0$ transitions (Soret region) of "bare" MgTBP. From Dr. Uzi Even

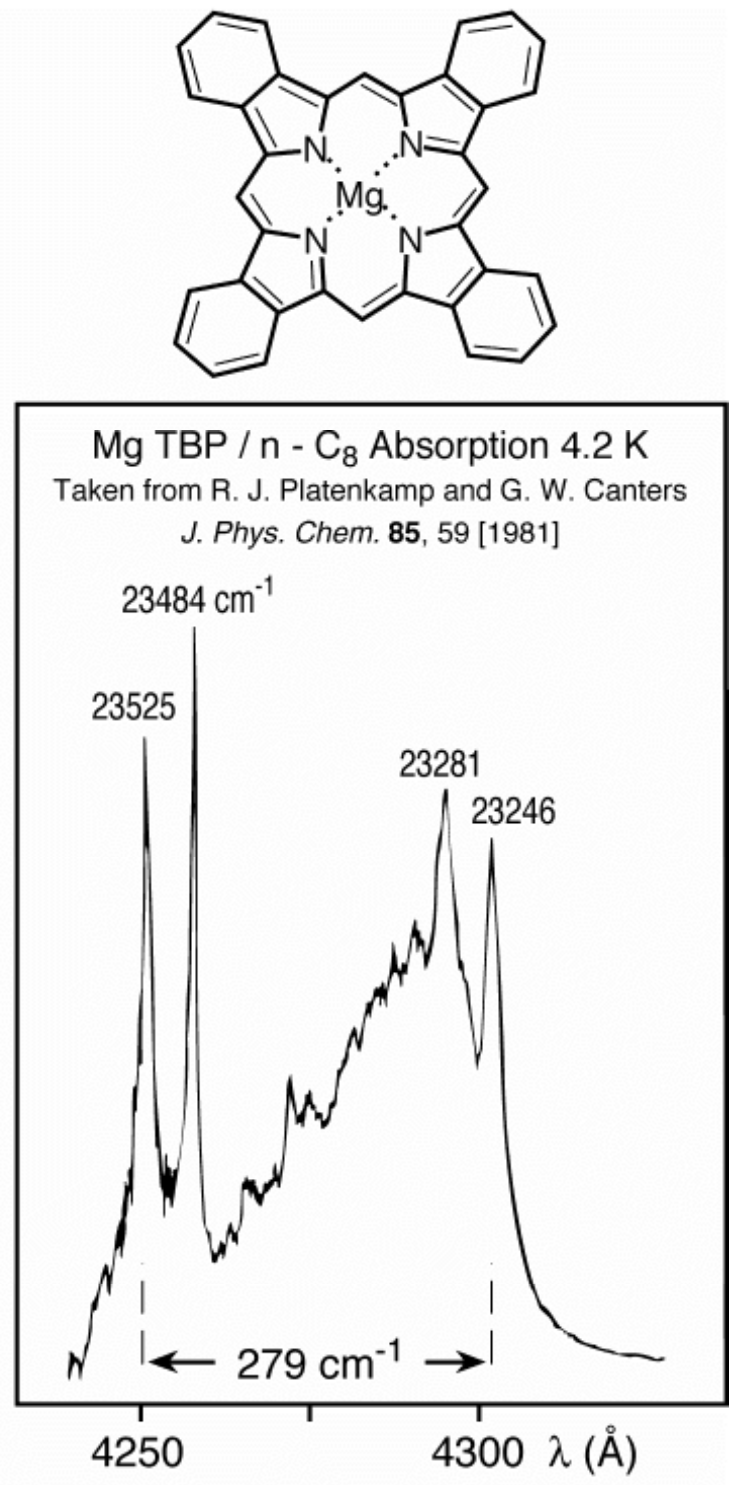

Figure 8b

The Soret band of MgTBP/n-$C_8$, at 4.2K [7].

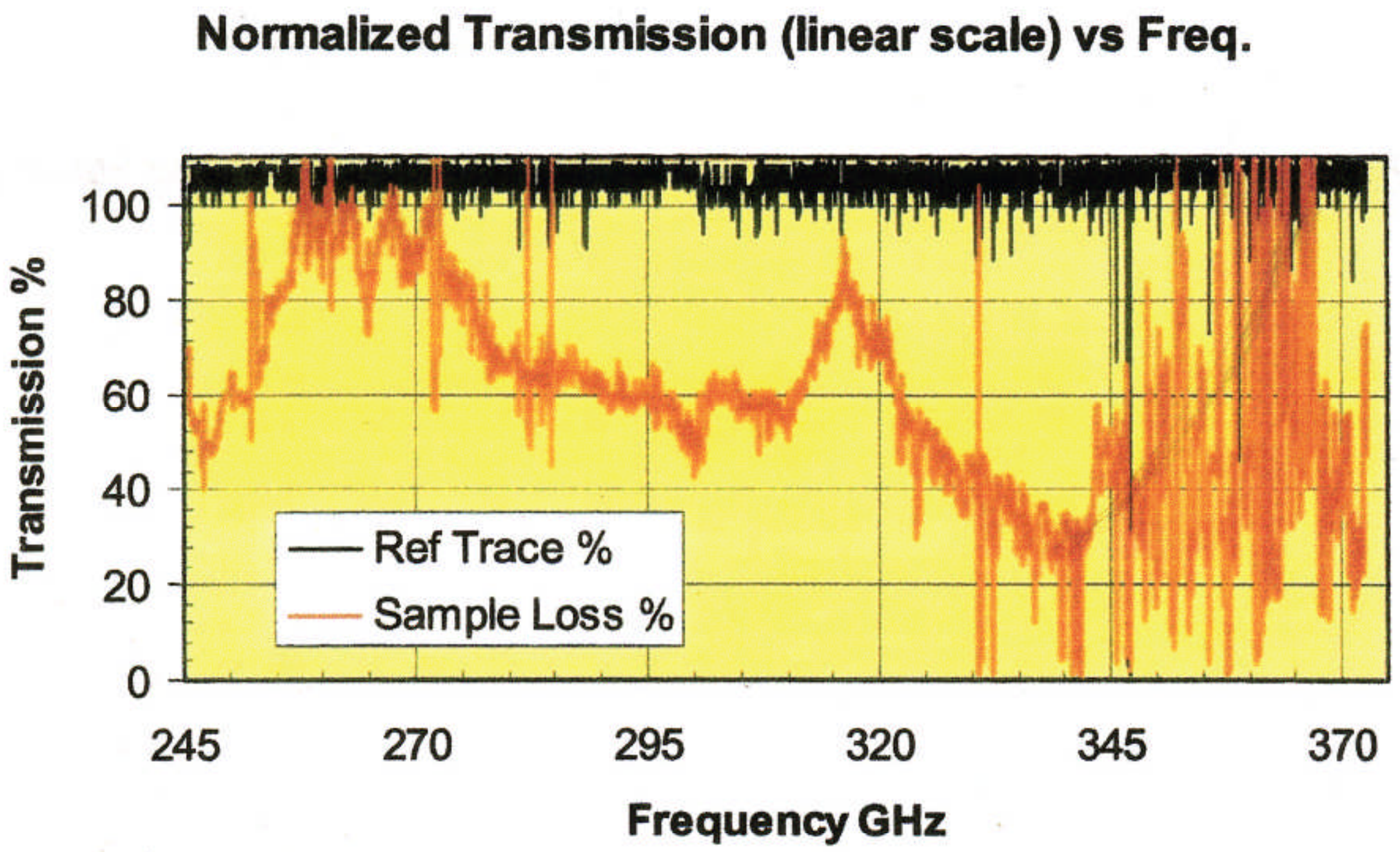

Figure 9. Microwave room temperature absorption measurement of MgTBP. The peak of its absorption is at 341 GHz. The latter value was confirmed by measuring its second harmonic at about 680 GHz. This graph was kindly provided by Dr. P.H. Siegel.

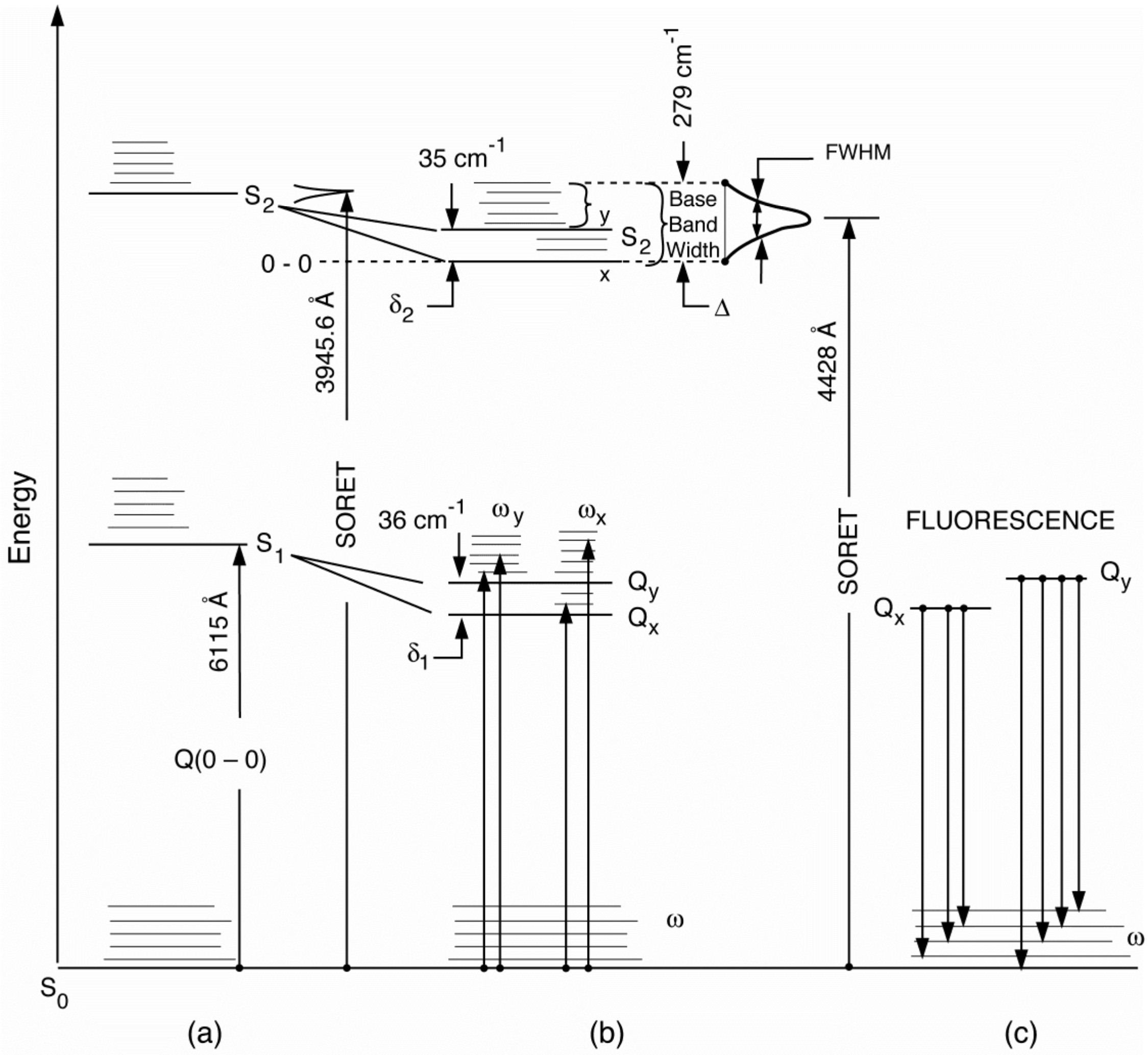

Figure 10. Schematic energy level diagram, showing vibronic transitions in absorption and in emission for MgTBP. Note: the transition from the "bare" state to the matrix and the splitting of the $S_1$ and $S_2$ states. Also note the Soret bands in the bare and in the matrix configuration.

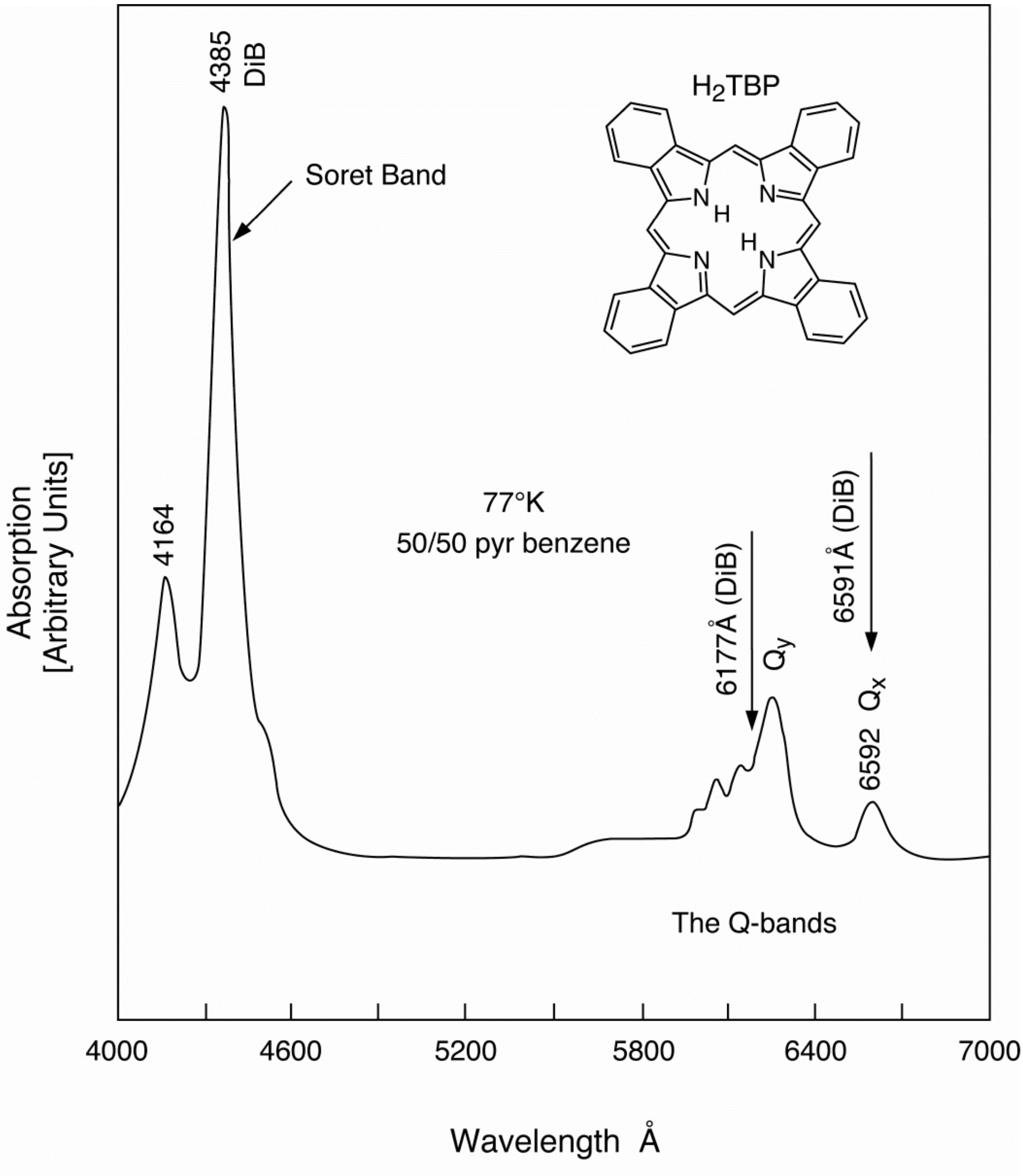

Figure 11. Non-Shpolskii lab absorption spectrum of $H_2TBP$ at 77K.
Note: the position of the 4385Å Soret band and the double Q-bands.
Author's lab data.

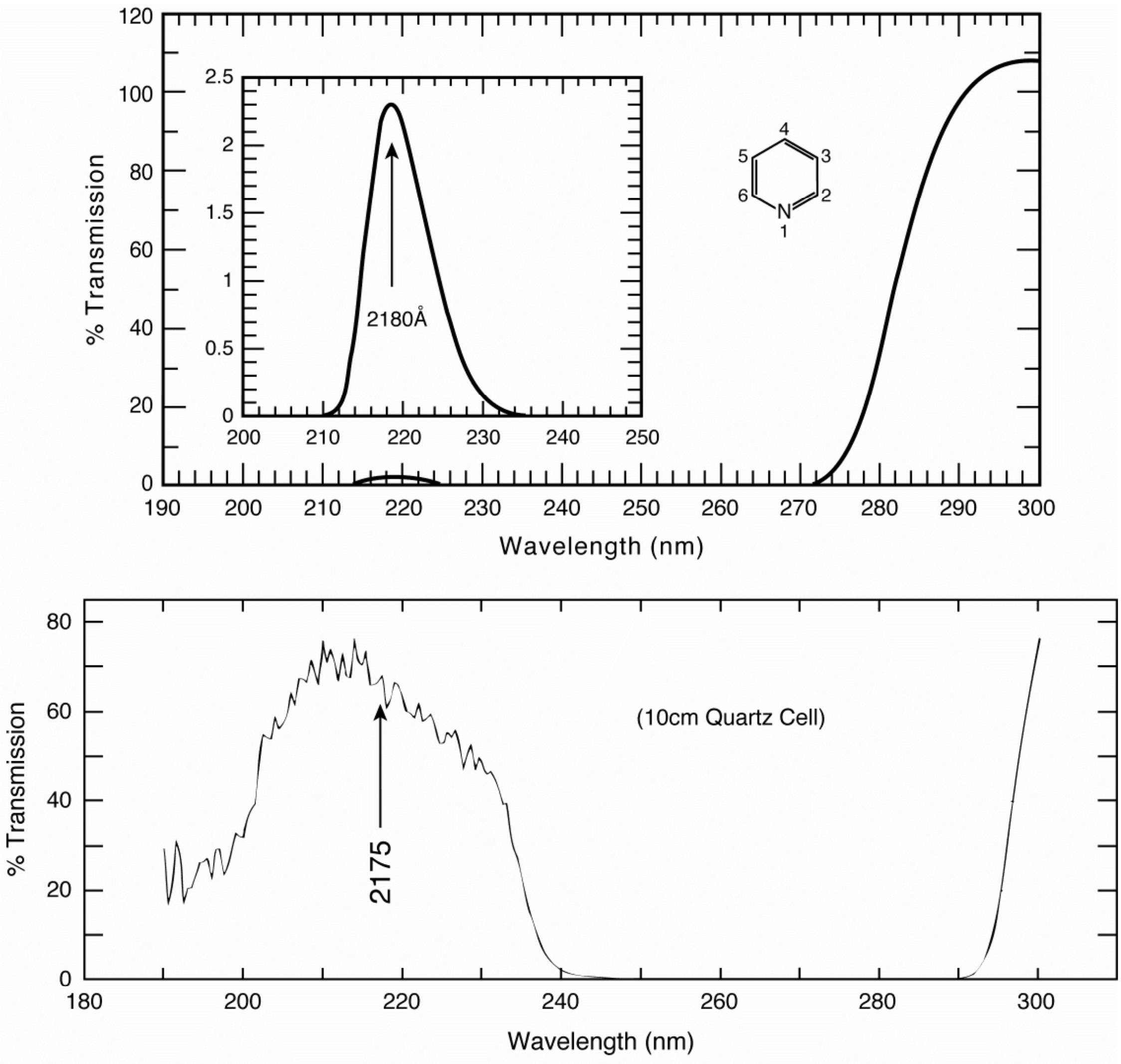

Figure 12. The UV transmission spectra of highly diluted pyridine in an octane solvent, at room temperature, at two different concentrations.
Author's lab data.

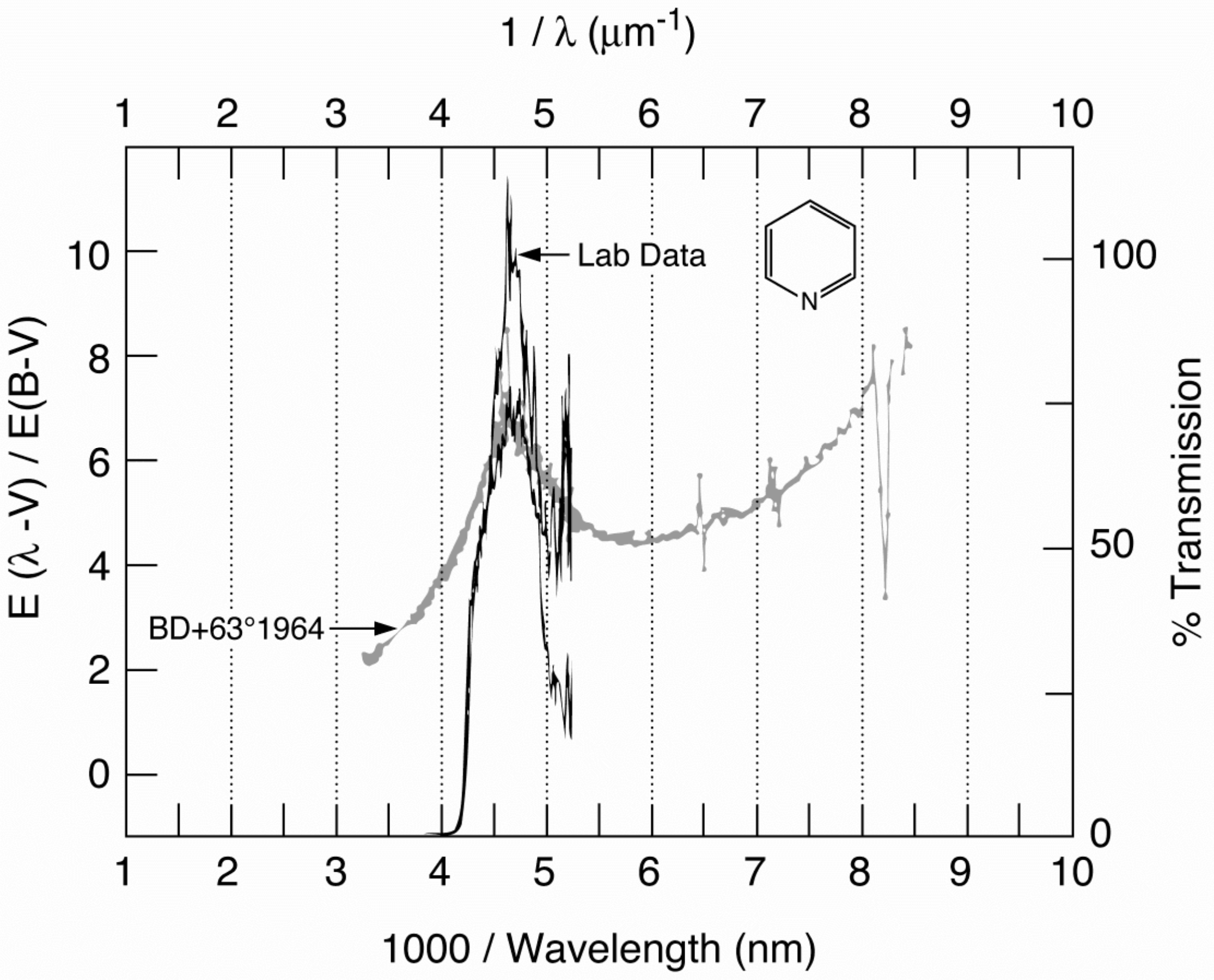

Figure 13. The superposition of the pyridine lab spectrum and a typical UV bump.

Author's lab data.

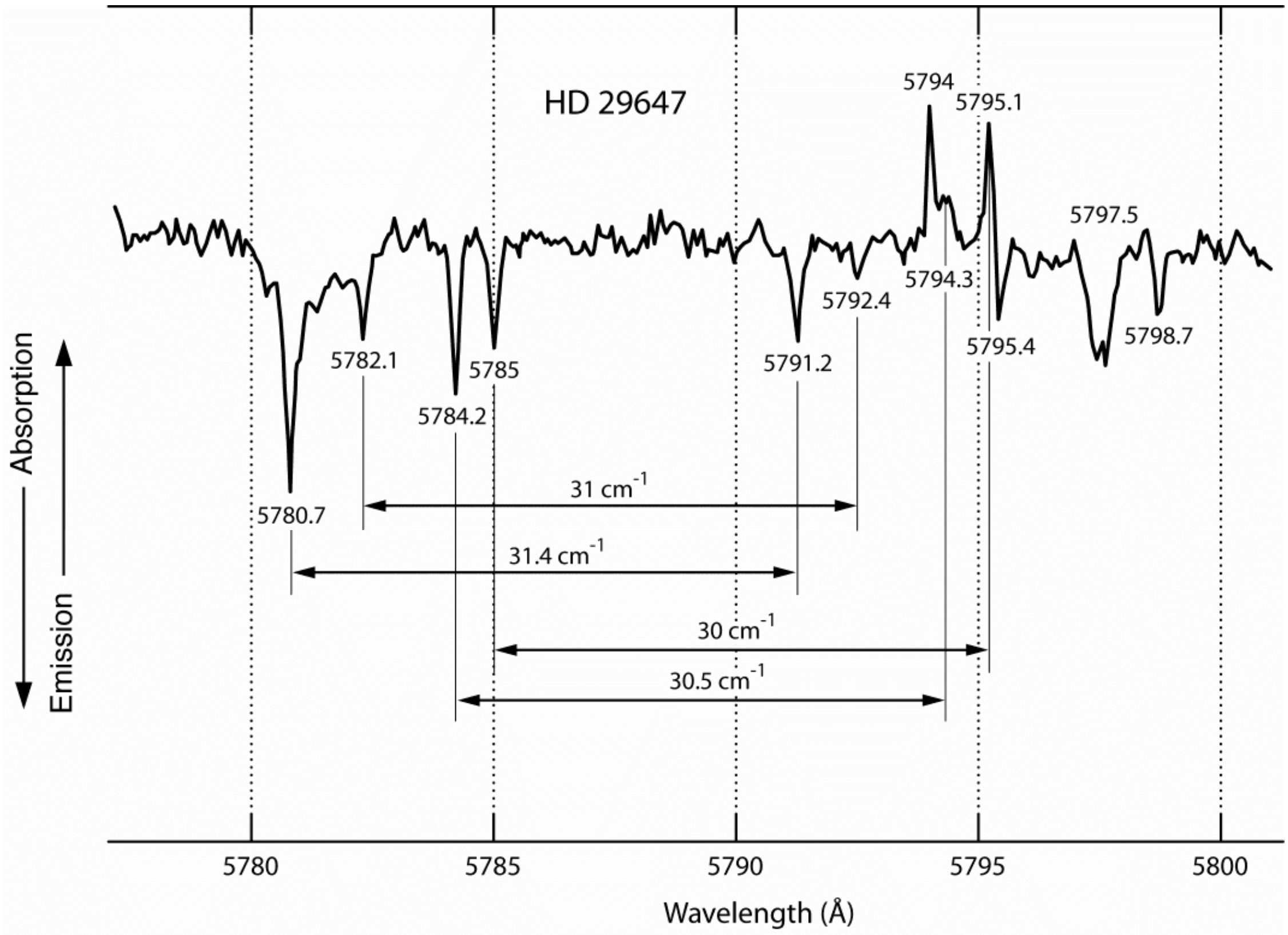

Figure 14. DIBs in HD29647. This graph was adapted from original data supplied by Dr. C.G. Seab. See text for interpretation.

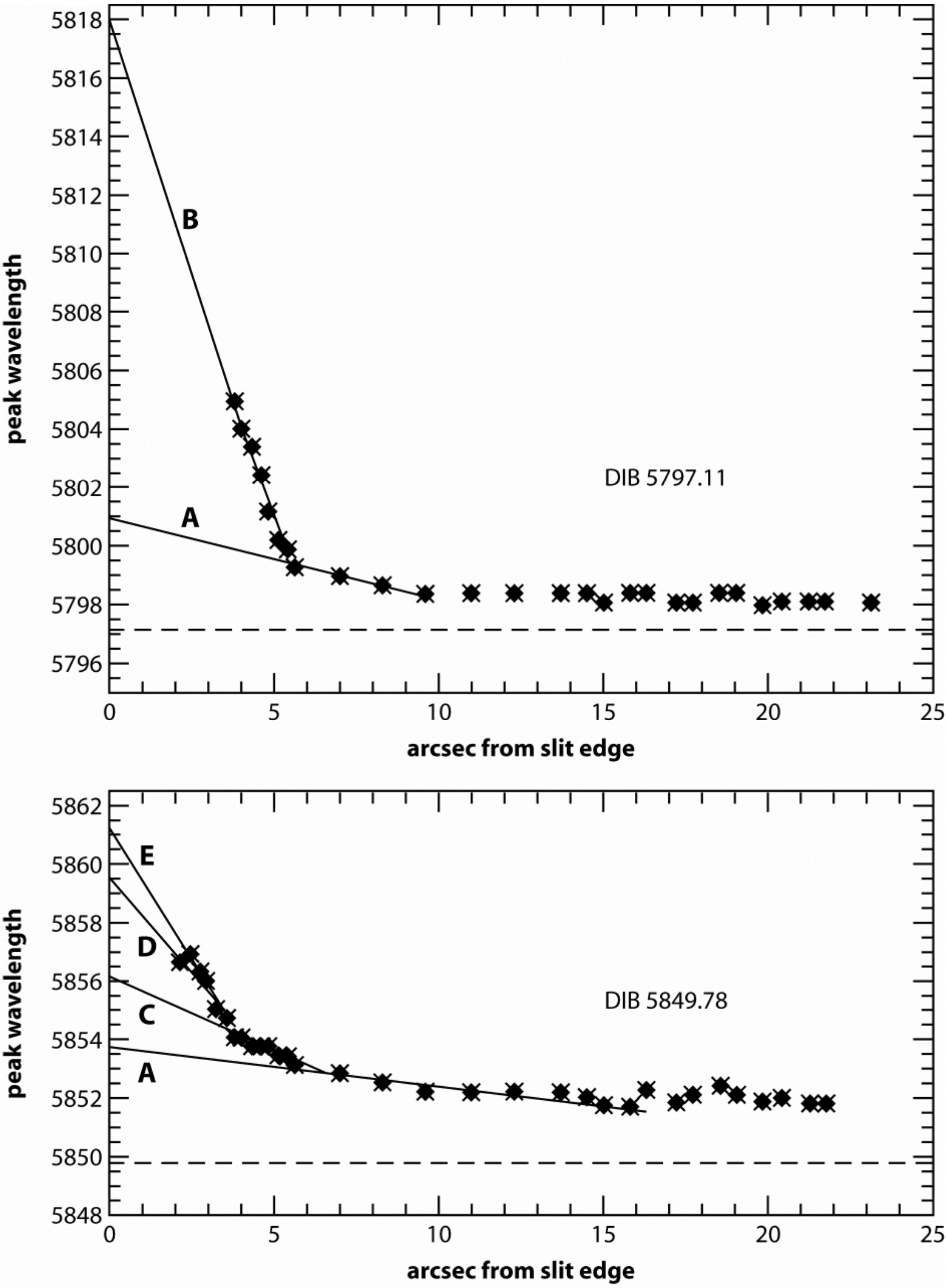

Figure 15. Adapted reproduction of Fig 9. from Van Winckel et al. [87]. The extrapolations from the data generate lines A-E (refer to text for implications)

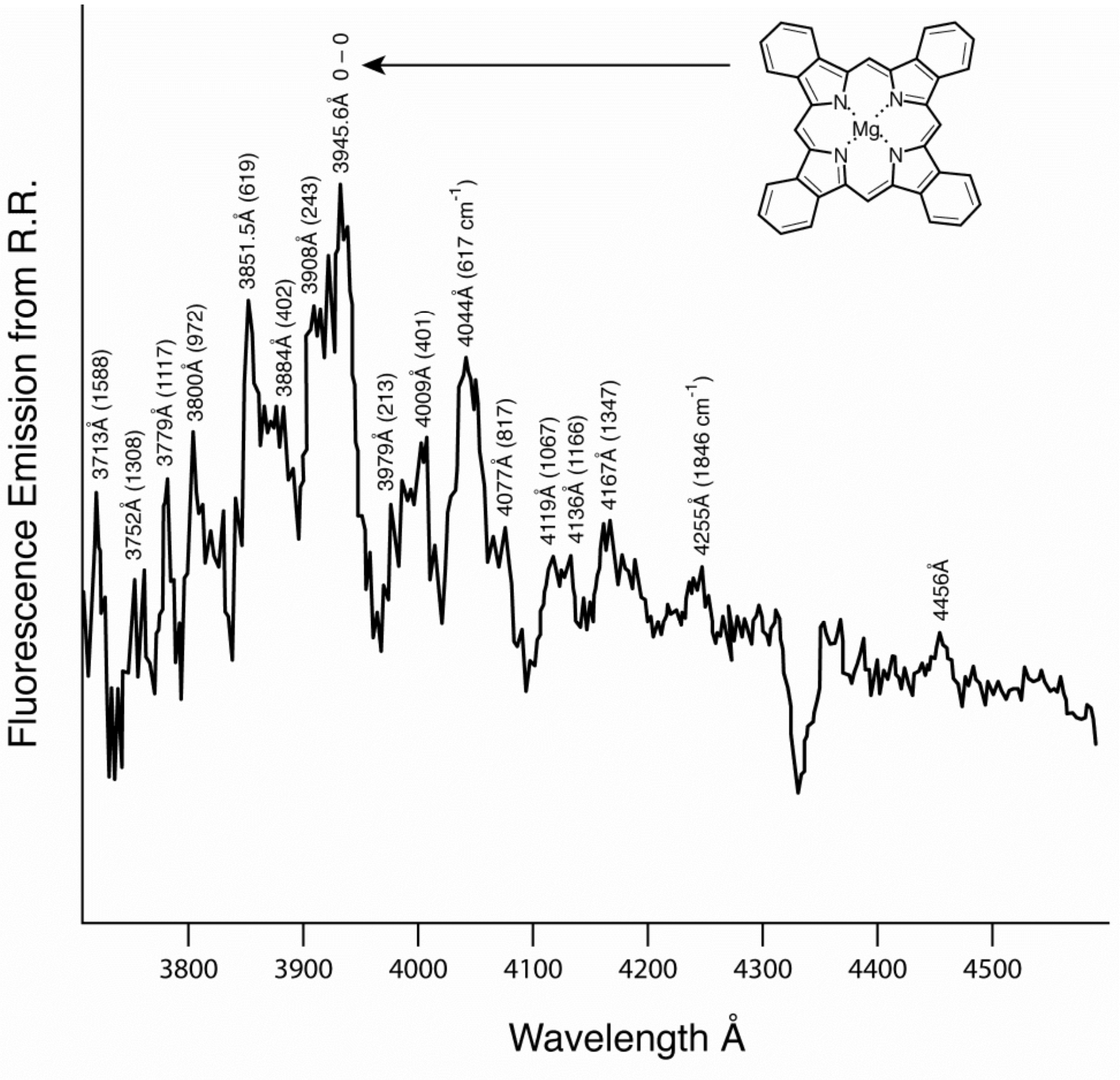

Figure 16. Adapted reproduction from Schmidt et al. [93]. The peaks indicate the fluorscence excitation spectrum associated with MgTBP. Assigned vibrational frequencies are indicated at each peak.

Peaks of published data measured by author. See text for implications.

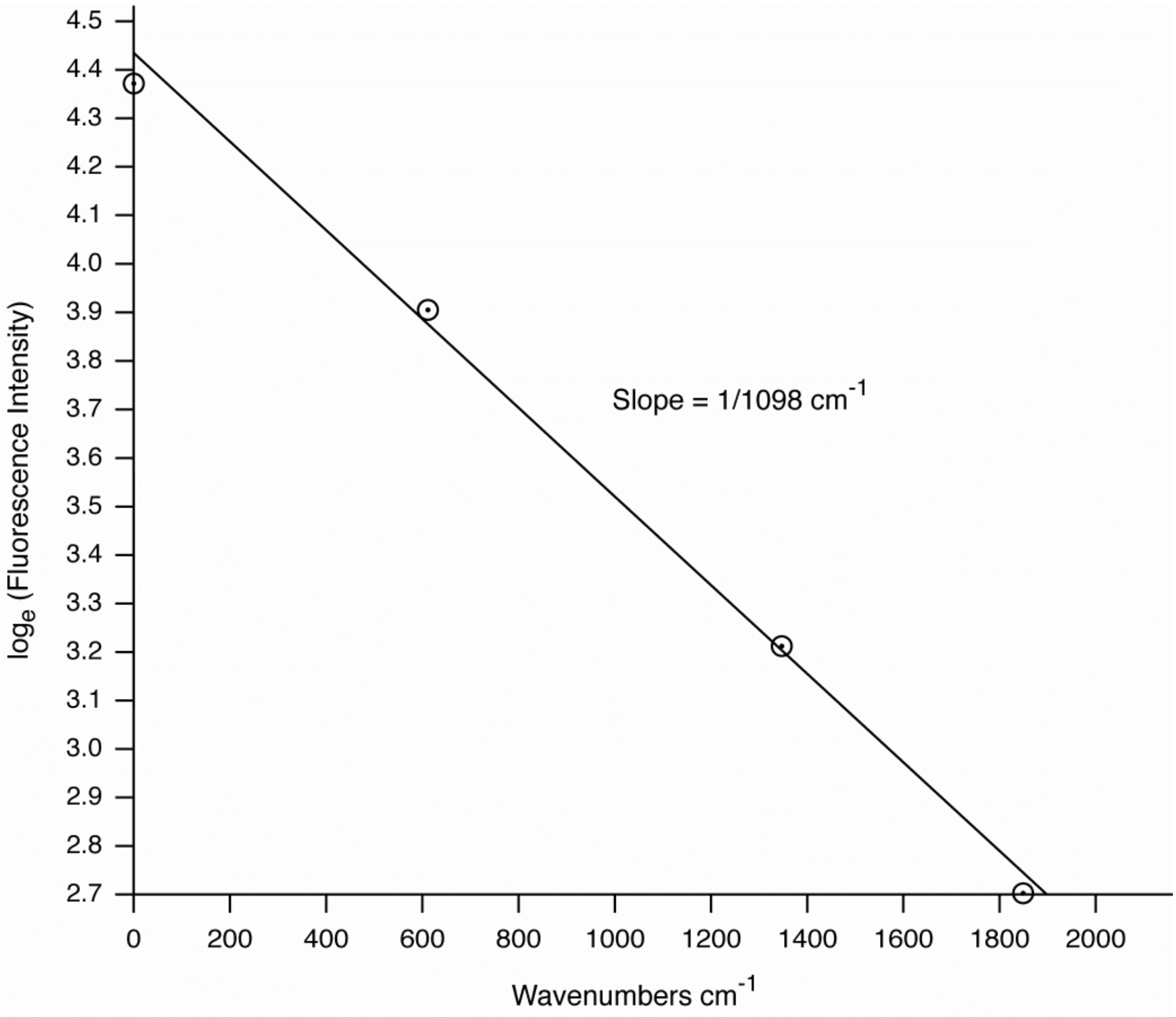

Figure 17. A plot of the log (intensity of fluorescence emission spectrum) vs. excitation energy above the ground state, utilizing the peak intensities of the hot bands (for $\lambda \geq 3945.6$Å).

Figure 1. Dipyridyl magnesium tetrabenzoporphyrin ($MgC_{46}H_{30}N_6$).

Figure 2. Energy (in $cm^{-1}$) level diagram (to scale) of "sites" of MgTBP in matrices of n-octane and octane/ pyridine mixtures for the $S_1$ electronic state. Sites I thru V are derived from author's data. Site VI is derived from [7].

Figure 3. The superposition of the MgTBP lab Soret band and the 4428Å DIB. Superpositioned (red) lab Soret band is from author's data.

Figure 4. Shpolskii matrix at 77K, showing a lab band at 6379Å, which encompasses DIBs λλ 6379, 6376, 6362 and 6353. Author's lab data superimposed with astronomical data.

Figure 5. The superposition of two spectral lab bands (using rescanned data by David Weinkle). Note: the Q-band in absorption as well as an emission spectrum. The Q-band illustrates the superposition of an inhomogeneously broadened Q-band, plus a narrow Shpolskii band in the 6284Å region. Spectral plates from author's data set.

Figure 6. Six individual, superimposed Shpolskii spectral scans. Note: the overlap of some of the individual spectral lines. Also shown are the relevant, corresponding DIBs. Spectral Data from author's data set.

Figure 7. Bare MgTBP LIF spectra. The inset displays the low frequency vibrations. This commissioned data was supplied by Dr. Don Brumbaugh

Figure 8a. The $S_2 \leftarrow S_0$ transition (Soret region) of "bare" MgTBP. From Dr. Uzi Even

Figure 8b. The Soret band of MgTBP/n-$C_8$, at 4.2K [7].

Figure 9. Microwave room temperature absorption measurement of MgTBP. The peak of its absorption is at 341 GHz. The latter value was confirmed by measuring its second harmonic at about 680 Ghz.  This graph was provided by Dr. P.H. Siegel.

Figure 10.  Schematic energy level diagram, showing vibronic transitions in absorption and in emission for MgTBP.  Note:  the transition from the "bare" state to the matrix and the splitting of the $S_1$ and the $S_2$ states.  Also note the Soret bands in the bare and in the matrix configuration.

Figure 11.  Non-Shpolskii lab absorption spectra of $H_2$TBP at 77K.  Note: the position of the 4385Å Soret band and the double Q-bands.  Author's lab data.

Figure 12.  The UV transmission spectrum of highly diluted pyridine in an octane solvent, at room temperature, at two different concentrations. Author's lab data.

Figure 13.  The superposition of the pyridine lab spectrum and a typical UV bump.  Author's lab data.

Figure 14.  DIBs in HD29647.   This graph was adapted from original data supplied by Dr. C.G. Seab.  See text for interpretation.

Figure 15.  Adapted reproduction of Fig. 9 from Van Winckel et al. [87].  The extrapolations from the data generate lines A-E (refer to text for implications).

Figure 16.  Adapted reproduction from Schmidt et al. [93].  The peaks indicate the fluorescence excitation spectrum associated with MgTBP.  Assigned vibrational frequencies are indicated at each peak.  Peaks of published data were measured by author.  See text for implications.

Figure 17.  A plot of the log (intensity of fluorescence emission spectrum) vs. excitation energy above the ground state, utilizing the peak intensities of the hot bands (for $\lambda \geq 3945.6$ Å).